\pdfoutput=1

\documentclass[]{aastex631}
\usepackage{amsmath,bm}
\usepackage{graphicx}
\usepackage{mathrsfs}
\usepackage{color}
\usepackage{url}
\usepackage{CJKutf8}
\usepackage{ulem}
\usepackage{wasysym}

\received{2022}
\revised{\today}
\accepted{\today}

\shorttitle{}
\shortauthors{
}

\begin{document}

\title{
The Lingering Death of Periodic Near-Sun Comet 323P/SOHO
}

\correspondingauthor{Man-To Hui}
\email{mthui@must.edu.mo, 
manto@hawaii.edu}

\author{\begin{CJK}{UTF8}{bsmi}Man-To Hui (許文韜)\end{CJK}}
\affiliation{State Key Laboratory of Lunar and Planetary Science, 
Macau University of Science and Technology, 
Avenida Wai Long, Taipa, Macau}

\author{David J. Tholen}
\affiliation{Institute for Astronomy, University of Hawai`i, 
2680 Woodlawn Drive, Honolulu, HI 96822, USA}

\author{Rainer Kracht}
\affiliation{Ostlandring 53, D-25335 Elmshorn, 
Schleswig-Holstein, Germany}

\author{\begin{CJK}{UTF8}{bsmi}Chan-Kao Chang (章展誥)\end{CJK}}
\affiliation{Institute of Astronomy and Astrophysics, Academia Sinica, 
No.1, Sec. 4, Roosevelt Rd, Taipei 10617, Taiwan}

\author{Paul A. Wiegert}
\affiliation{Department of Physics and Astronomy, 
The University of Western Ontario, London, Ontario N6A 3K7, Canada}
\affiliation{Institute for Earth and Space Exploration, 
The University of Western Ontario, London, Ontario N6A 3K7, Canada}

\author{\begin{CJK}{UTF8}{bsmi}Quan-Zhi Ye (葉泉志)\end{CJK}}
\affiliation{Department of Astronomy,
University of Maryland, College Park, MD 20742, USA}

\author{Max Mutchler}
\affiliation{Space Telescope Science Institute, Baltimore, 
3700 San Martin Drive, Baltimore, MD 21218, USA}


\begin{abstract}

We observed near-Sun comet 323P/SOHO for the first time using ground and space telescopes. In late December 2020, the object was recovered at Subaru showing no cometary features on its way to perihelion. However, in our postperihelion observations it developed a long narrow tail mimicking a disintegrated comet. The ejecta, comprised of at least mm-sized dust with power-law size distribution index $3.2 \pm 0.2$, was impulsively produced shortly after the perihelion passage, during which $\ga$0.1-10\% of the nucleus mass was shed due to excessive thermal stress and rotational disruption. Two fragments of $\sim$20 m in radius (assuming a geometric albedo of 0.15) were seen in HST observations from early March 2021. The nucleus, with an effective radius of $86 \pm 3$ m (the same albedo assumed) and an aspect ratio of $\sim$0.7, has a rotation period of 0.522 hr, which is the shortest for known comets in the solar system and implies cohesive strength $\ga$10-100 Pa in the interior. The colour of the object was freakish, and how it changed temporally has never been previously observed. Using our astrometry, we found a strong nongravitational effect following a heliocentric dependency of $r_{\rm H}^{-8.5}$ in the transverse motion of the object. Our N-body integration reveals that 323P has a likelihood of 99.7\% to collide with the Sun in the next two millennia driven by the $\nu_6$ secular resonance.

\end{abstract}


\section{Introduction}
\label{intro}

The near-Sun population of small bodies is a group of comets and asteroids with perihelion distances smaller than that of Mercury \citep[$q \la 0.31$ au;][]{2018SSRv..214...20J}. They are predicted to be common dynamical end states of main-belt asteroids or short-period comets that were gravitationally scattered by major planets and/or diverted by nongravitational forces \citep{1992A&A...257..315B, 1994Natur.371..314F, 1997Sci...277..197G, 2012Icar..217..355G}. Typical dynamical lifetimes of near-Sun objects are only $\la$10 Myr due to frequent crossing of the orbits of terrestrial planets \citep{1997Sci...277..197G}. \\

It is known that the observed number of objects in the near-Sun population is much scarcer than dynamical models \citep[e.g.,][]{1994Natur.371..314F, 2012Icar..217..355G} predict. While the unfavourable observability of near-Sun objects, which get bright enough only when near perihelia at extremely small solar elongations, certainly plays a role, their thermal destruction by which they fragment into millimeter-sized particles is no less important \citep{2016Natur.530..303G, 2020AJ....159..143W}. To date there have been no direct good-quality observations showing the fragmentation process. Evidence that appears to support the destruction hypothesis all comes from ultra-low resolution and low-sensitivity observations from solar probes. They are usually the only dataset available for studies of the near-Sun population, resulting in an extremely poor understanding of this group of objects and their fragmentation process, as exemplified by arguably the most famous case (3200) Phaethon \citep{2010AJ....140.1519J,2013AJ....145..154L,2017AJ....153...23H}. \\

323P/SOHO is a periodic near-Sun comet discovered by the Solar and Heliospheric Observatory (SOHO) in 1999 and is not linked to any of the identified near-Sun dynamical groups \citep{2013Icar..226.1350L}. Despite no clear detection of cometary features, the anomalous brightening around perihelion indicates its repetitive activity thereabouts, unlike Kreutz sungrazing comets, many of which exhibit visible cometary features even in the low-resolution SOHO images \citep{2010AJ....139..926K}. The comet orbits around the Sun every $\sim$4.2 yr and passed perihelion on 2021 January 17 at a perihelion distance of $q = 0.04$ au (or $\sim$8.4 $R_{\odot}$, where $R_{\odot} = 6.96 \times 10^{5}$ km is the solar radius). It had never been observed from the ground prior to our observations. As such, 323P becomes only the second SOHO-discovered periodic comet observed by non-solar observatories after 322P/SOHO \citep{2016ApJ...823L...6K}. In this paper, we detail observations of 323P in Section \ref{sec_obs}, present results and discussion in Sections \ref{sec_res} and \ref{sec_disc}, respectively, and conclude in Section \ref{sec_sum}. In a nutshell, we observed a periodic near-Sun comet and its mass loss in great detail for the very first time. \\

\section{Observation}
\label{sec_obs}

We observed 323P using ground and space telescopes, including the Canada-France-Hawaii Telescope (CFHT), Gemini North (GN), Hubble Space Telescope (HST), Lowell Discovery Telescope (LDT), and Subaru, between 2020 December and 2021 March, covering both the inbound and outbound legs of the comet's orbit (Table \ref{tab:vgeo}). The perihelion passage of the comet was unable to be monitored from the aforementioned telescopes due to the extremely unfavourable observing geometry but was exclusively visible from SOHO. However, the SOHO observations were all taken by cameras onboard having significantly worse sensitivity and resolution, and were therefore only included for astrometry. During our observing campaign, 323P passed perigee on 2021 February 07 at an unremarkable close-approach distance of 0.440 au. In the following we describe our observations from each telescope in detail.

\begin{deluxetable*}{lcccccccccccc}
\tabletypesize{\footnotesize}
\tablecaption{Observing Geometry of 323P/SOHO
\label{tab:vgeo}}
\tablewidth{0pt}
\tablehead{ 
\colhead{Date (UT)} & \colhead{Telescope\tablenotemark{a}} & \colhead{Filter} & \colhead{\#\tablenotemark{b}} & \colhead{$t_{\rm exp}$ (s)\tablenotemark{c}} & \colhead{$r_{\rm H}$ (au)\tablenotemark{d}}  & 
\colhead{$\it \Delta$ (au)\tablenotemark{e}} & \colhead{$\alpha$ (\degr)\tablenotemark{f}} & 
\colhead{$\varepsilon$ (\degr)\tablenotemark{g}} &
\colhead{$\theta$ (\degr)\tablenotemark{h}} &
\colhead{$\theta_{-\odot}$ (\degr)\tablenotemark{i}} &
\colhead{$\theta_{-{\bf V}}$ (\degr)\tablenotemark{j}} &
\colhead{$\psi$ (\degr)\tablenotemark{k}}
}
\startdata
2020 Dec 21 & Subaru & $r$ & 4 & 170 & 0.914 & 1.078 & 58.5 & 52.4 & 201.7 & 288.2 & 289.3 & $-4.0$ \\
\hline
2021 Feb 06 & CFHT & $gri$ & 3 & 60 & 0.737 & 0.442 & 111.1 & 44.2 & 155.3 & 61.6 & 241.2 & $-1.4$ \\
 \hline
2021 Feb 07 & CFHT & $gri$ & 3 & 120 & 0.762 & 0.440 & 107.2 & 47.6 & 155.7 & 62.1 & 241.7 & $-1.2$ \\
 \hline
2021 Feb 08 & CFHT & $gri$ & 3 & 120 & 0.787 & 0.441 & 103.2 & 51.0 & 156.2 & 62.7 & 242.4 & $-1.0$ \\
 \hline
2021 Feb 10 & CFHT & $gri$ & 3 & 120 & 0.835 & 0.450 & 95.6 & 57.4 & 157.0 & 64.2 & 244.0 & $-0.6$ \\
 \hline
2021 Feb 11 & CFHT & $gri$ & 4 & 120 & 0.859 & 0.458 & 92.0 & 60.4 & 157.4 & 65.0 & 244.9 & $-0.4$ \\
 \hline
2021 Feb 12 & CFHT & $gri$ & 3 & 120 & 0.882 & 0.467 & 88.4 & 63.3 & 157.8 & 65.9 & 245.9 & $-0.2$ \\
 \hline
2021 Feb 13 & CFHT & $gri$ & 3 & 120 & 0.905 & 0.479 & 85.1 & 66.0 & 158.1 & 66.9 & 246.9 & $+0.0$ \\
 \hline
\vspace{-0.1cm} &  & $g'$ & 5 & 150 & &  &  &  &&& & \\ \vspace{-0.1cm}
2021 Feb 13 & GN & $r'$ & 6 & 140 & 0.906 & 0.480 & 85.0 & 66.0 & 158.2 & 66.9 & 246.9 & $+0.1$ \\
 &  & $i'$ & 2 & 135 &  &  &  &  &&& & \\
 \hline
\vspace{-0.1cm} &  & $r$ & 4 & & &  &  &  &&& & \\ \vspace{-0.1cm}
2021 Feb 16 & LDT & & & 180 & 0.968 & 0.521 & 76.6 & 72.5 & 159.0 & 69.6 & 249.8 & $+0.6$ \\
 &  & $VR$ & 5 & &  &  &  &  &&& & \\
 \hline
2021 Feb 17 & CFHT & $gri$ & 2 & 120 & 0.993 & 0.541 & 73.7 & 74.6 & 159.4 & 70.7 & 250.9 & $+0.7$ \\
 \hline
2021 Mar 02 & HST & F350LP & 5 & 405 & 1.255 & 0.844 & 52.0 & 86.0 & 162.3 & 80.2 & 260.8 & $+1.9$ \\
\hline
2021 Mar 03 & LDT & $VR$ & 18 & 180 & 1.269 & 0.862 & 51.2 & 86.1 & 162.4 & 80.6 & 261.1 & $+1.9$ \\
 \hline
\vspace{-0.1cm} &  & $g'$ & 9 & 150 & &  &  &  &&& & \\ \vspace{-0.1cm}
2021 Mar 03 & GN & $r'$ & 10 & 140 & 1.270 & 0.864 & 51.1 & 86.1 & 162.4 & 80.6 & 261.2 & $+1.9$ \\
 &  & $i'$ & 7 & 135 &  &  &  &  &&& & \\
 \hline
2021 Mar 03 & HST & F350LP & 5 & 405 & 1.271 & 0.866 & 51.1 & 86.1 & 162.4 & 80.6 & 261.2 & $+1.9$ \\
\hline
2021 Mar 22 & HST & F350LP & 5 & 405 & 1.592 & 1.378 & 38.4 & 82.5 & 165.0 & 87.1 & 268.0 & $+2.4$ \\
\hline
2021 Mar 26 & HST & F350LP & 5 & 405 & 1.655 & 1.490 & 36.5 & 80.8 & 165.4 & 88.0 & 269.0 & $+2.4$ \\
\enddata
\tablenotetext{a}{CFHT = 3.6 m Canada-France-Hawaii Telescope, GN = 8.1 m Gemini North telescope, HST = 2.4 m Hubble Space Telescope, LDT = 4.3 m Lowell Discovery Telescope, Subaru = 8.2 m Subaru telescope.}
\tablenotetext{b}{Number of useful exposures, where the comet is uninvolved with background sources.}
\tablenotetext{c}{Individual exposure time.}
\tablenotetext{d}{Heliocentric distance.}
\tablenotetext{e}{Observer-centric distance.}
\tablenotetext{f}{Phase angle (Sun-comet-observer).}
\tablenotetext{g}{Solar elongation (Sun-observer-comet).}
\tablenotetext{h}{True anomaly.}
\tablenotetext{i}{Position angle of projected antisolar direction.}
\tablenotetext{j}{Position angle of projected negative heliocentric velocity of the comet.}
\tablenotetext{k}{Observer to comet's orbital plane angle with vertex at the comet. Negative values indicate observer below the orbital plane of the comet.}

\end{deluxetable*}

\begin{figure}
\epsscale{1.0}
\begin{center}
\plotone{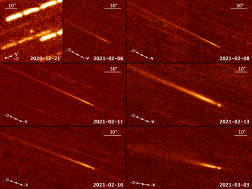}
\caption{
The appearance of 323P in our Subaru (2020 December 21, $r$-band), CFHT (2021 February 6, 8, and 11, $gri$ filter), LDT (2021 February 16, $r$ and $VR$ filters), and GN (2021 February 13 and March 3, $r$-band) images from late December 2020 to early March 2021. Each panel is median combined from individual exposures in the aforementioned corresponding filters from the same observing night, with registration on the comet and background sources masked out, except for the Subaru panel, which is an average from the background-offset image sequence and then convolved with a Gaussian of 2 pixels in FWHM so as to cosmetically suppress background noise. The antisolar direction ($-\odot$) and the negative heliocentric velocity of the comet projected onto the sky plane ($-{\bf V}$) are shown as the arrows. Corresponding scale bars are given in each panel. Equatorial J2000 north is up and east is left.
\label{fig:grnd_obs}
} 
\end{center} 
\end{figure}

\begin{figure}
\epsscale{1.0}
\begin{center}
\plotone{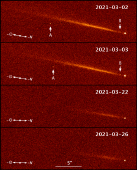}
\caption{
Same as Figure \ref{fig:grnd_obs}, but in HST/WFC3 median combined images from March 2021. Two fragments, labelled as ``A" and ``B", respectively, and pointed out by the two arrows, can be seen in the upper two panels. A common scale bar of 5\arcsec~is shown at the bottom. Equatorial J2000 north is up and east is left.
\label{fig:HST_obs}
} 
\end{center} 
\end{figure}

\subsection{Subaru}

The 8.2 m Subaru telescope equipped with the Hyper Suprime-Cam \citep[HSC;][]{2018PASJ...70S...1M} at the prime focus was employed to search for 323P on its way to perihelion on 2020 December 21 (Programme S20B-UH014-A). At that time the ephemeris uncertainty of the comet was so enormous that the $3\sigma$ uncertainty ellipse had major and minor axes of $\sim$16\arcmin~and 9\arcmin, respectively. This was because the comet had only been observed in low-resolution (pixel scale $\ga$10\arcsec) SOHO images, resulting in a highly uncertain orbital solution. However, the HSC, which is mosaicked from an array of 104 main $2048 \times 4096$ pixel science CCDs and has a gigantic FOV of $\sim$1\fdg5 in diameter with a pixel scale of 0\farcs17, facilitated us to effectively cover the search region and find the target. Six $r$-band images were taken consecutively tracking at the ephemeris nonsidereal motion rate of the comet ($\sim$4\farcm3 hr$^{-1}$) and dithered between each exposure to reduce the risk that the comet could accidentally fall into the CCD chip gaps. Because of the high airmass of the observation ($\sim$2), all of the images suffered from bad seeing ($\sim$1\farcs7).

\subsection{Canada-France-Hawaii Telescope}

A large number of our observations were taken with the MegaCam prime focus imager \citep{2003SPIE.4841...72B} using the broadband $gri$ filter at the 3.6 m CFHT atop Maunakea, Hawai`i. The device is a mosaic of 40 CCD chips, each having a common angular field-of-view (FOV) of $6\farcm4 \times 14\farcm4$ and an image scale of 0\farcs374 pixel$^{-1}$ in the $2 \times 2$ binning mode, which was performed during the image calibration process. The telescope was tracked at the nonsidereal apparent motion rate of 323P, which was nontrivial ($\sim$7\arcmin-11\arcmin~hr$^{-1}$), resulting in background stars obviously trailed. We estimated the seeing by measuring the full width at half maximum (FWHM) of star trails in the cross-track direction to be in a range of 0\farcs6 and 1\farcs8, varying from night to night. Images from each night were taken continuously with dithering between each exposure so as to mitigate CCD defects.

\subsection{Gemini North}

We obtained $g'$, $r'$, and $i'$-band observations of 323P using the Gemini Multi-Object Spectrograph at the 8.1 m Gemini North telescope \citep[GMOS-N;][]{2004PASP..116..425H}, also on the summit of Maunakea on 2021 February 13, when Earth was near the orbital plane of 323P, and March 3 (Programme GN-2021A-DD-201). The GMOS-N Hamamatsu CCD array consists of three CCDs covering an overall FOV of $5\farcm5 \times 5\farcm5$. To improve the observing efficiency the detector was read out with pixels $2 \times 2$ binned on chip, which provided us with an on-sky angular sampling of 0\farcs16 pixel$^{-1}$. The telescope also followed the apparent motion of the comet unguided. The seeing was not optimal, varying between $\sim$0\farcs8 and 1\farcs9 during the Gemini observations. Dithering between individual exposures was also performed.

\subsection{Hubble Space Telescope}

We employed the 2.4 m HST to observe 323P in four Director's Discretionary (DD) orbits (Programme GO/DD 16496). The observations were executed using the Wide Field Camera 3 \citep[WFC3;][]{Dressel2021}, which is comprised of two $2048 \times 4096$ pixel CCDs with an image scale of 0\farcs04 pixel$^{-1}$ in the UVIS channel, rendering a FOV of $2\farcm7 \times 2\farcm7$. For maximum sensitivity, the long-pass F350LP filter, which has an effective wavelength of 5846 \AA~and an FWHM of 4758 \AA, was used. In each DD orbit, five images were obtained, with dithering executed between the third and fourth exposures to help reduce effects from the inter-chip gap and bad pixels. As the observations followed the apparent motion of 323P and the HST orbited around Earth, background sources are all apparently trailed and slightly curved. Unfortunately, the HST observations were interrupted due to a series of malfunctions of the telescope that shut WFC3 down after completion of the first two DD visits in early March 2021. 

\subsection{Lowell Discovery Telescope}

We also observed 323P on 2021 February 16 and March 3 through the $r$-band and $VR$ filters using the the Large Monolithic Imager \citep[LMI;][]{2013AAS...22134502M} on the 4.3 m LDT tracking nonsidereally. The images have a square FOV of $12\farcm3 \times 12\farcm3$ and a pixel scale of 0\farcs36 after an on-chip $3 \times 3$ binning. The seeing was highly variable between $\sim$0\farcs9 and 2\farcs3, measured from the FWHM of background star trails in the cross-track direction in images. The first night at LDT witnessed influences from intermittent clouds at the beginning, and therefore we had to discard the first three exposures of the comet, in which field stars are no better than marginally seen.

\section{Results}
\label{sec_res}

All of the obtained images on 323P were calibrated with corresponding bias and flat frames. We managed to recover 323P from two HSC CCD chips in four of the Subaru images, in which the comet appeared asteroidal with an FWHM of $\sim$1\farcs7, indistinguishable from the seeing disc during the observation. However, after the perihelion passage, the morphology of the comet became drastically different. In our earliest postperihelion observation from CFHT on 2021 February 6, the comet has developed a long narrow tail, which was pinched off from the barely visible optocentre, mimicking a disintegrated comet. Thereafter the optocentre became progressively more obvious in our observations while the long narrow tail (at least $\ga$5\arcmin~in length early on) persisted throughout the remainder of the observing campaign (Figure \ref{fig:grnd_obs}).

The general morphology of 323P observed in the HST/WFC3 is basically the same as seen from the ground telescopes. However, we managed to robustly identify two fragments of 323P both approximately in the tailward direction of the comet in the first two visits from 2021 March 2 and 3 (Figure \ref{fig:HST_obs}). Here, we term the component that was apparently further from the primary Fragment A, and the other one Fragment B. Except in one single exposure from the first HST visit due to a cosmic ray hit, Fragment A was clearly visible in the other nine individual images from the first two HST visits. In comparison, Fragment B was only marginally discernible in individual images from the first visit but became more obvious in the second visit. Unfortunately, both of the fragments were lost in the last two HST visits in late March 2021, after the hiatus due to the malfunction of the telescope. We could not identify additional fragments of 323P in the HST/WFC3 images.

\subsection{Photometry}
\label{ssec_phot}

\begin{figure}
\epsscale{1.}
\begin{center}
\plotone{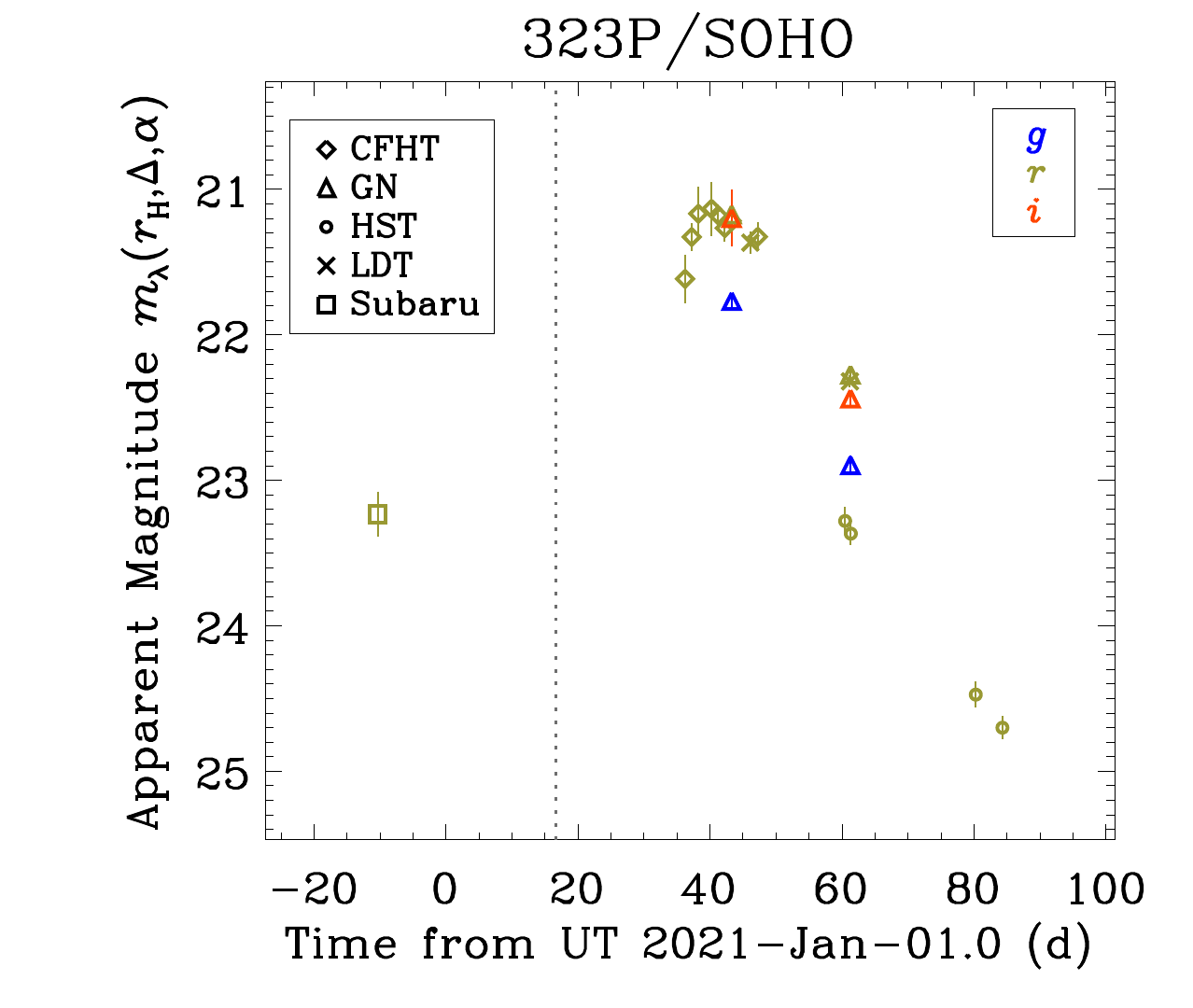}
\caption{
Temporal variation of the nightly mean apparent magnitude of 323P/SOHO measured with a circular aperture of fixed radius 2000 km projected at the distance of the comet except for the HST data points, which are photometric measurements of the nucleus. The perihelion epoch of the comet (TDB 2021 January 17.6) is marked by the vertical dotted line. As indicated in the label, data points from different telescopes are discriminated according to symbols, and the reduced bandpasses are colour coded.
\label{fig:mapp}
} 
\end{center} 
\end{figure}

We performed circular aperture photometry for the nucleus of 323P in our ground and space observations. To compensate for potential effects from the varying seeing in the ground-based images as well as to minimise contamination from the dust ejecta, we picked a fixed seeing aperture of 1$\times$ seeing FWHM in radius centred at the optocentre of the comet. The sky background was computed using annuli with inner and outer radii of 4$\times$ and 8$\times$ seeing FWHM, respectively, from the optocentre. Varying the annulus does not affect the obtained sky background value beyond the associated uncertainty, which was propagated from the Poisson statistics. The nucleus flux was then calculated by applying correction for an aperture effect assuming a bidimensional Gaussian brightness profile for the nucleus. Using increasingly larger sizes of fixed seeing apertures results in a systematically brighter nucleus flux, implying that the signal of the nucleus is progressively obscured by the surrounding dust ejecta. 

As for the HST observations, which are free from any atmospheric seeing effects, we simply employed a circular aperture of 2 pixels in radius, consistent with the Nyquist sampling of the HST/WFC3 images. The sky background was calculated using annuli having radii from 10 to 20 pixels from the optocentre. We then conducted aperture correction by applying the same photometric aperture on a point-spread function (PSF) for WFC3 images in the F350LP filter generated by {\tt TinyTim} \citep{2011SPIE.8127E..0JK}. 

In order to compensate for the varying observing geometry and to better characterise the surrounding ejecta of the comet, we adopted a series of circular apertures having fixed radii from 1000 to 3000 km in 500 km increments projected at the distance of the comet. For HST observations, due to their far superior angular resolution and sensitivity, we could not apply the same apertures, otherwise there would always be trails of background sources corrupting the photometric measurements. Given the observed intricate morphology of the dust ejecta, it is extremely difficult to apply any meaningful aperture correction for the HST photometry so as to be compatible with the photometry from ground telescopes. Therefore, we did not use aperture photometry to characterise the dust ejecta of the comet in the HST observations.

The measured fluxes can be converted to apparent magnitude with the corresponding image zero-points. However, only the HST/WFC3 images have a precisely determined image zero-point, which we obtained from the WFC3 UVIS Imaging Exposure Time Calculator. Thus, before we could perform the conversion, we had to determine image zero-points for the ground-based data. Since all of the observations were tracked at the apparent motion rate of the comet, background stars are significantly trailed, making simple centroiding algorithms for point sources inapplicable. To overcome this issue, we utilised a specific algorithm suited for measuring trailed images, where a source model is a trapezoid in the along-track direction and a Gaussian in the cross-track direction. The algorithm has a decades-long track record of providing high-quality astrometry and photometry of asteroids published by the Minor Planet Center. For each background trail, we performed the least-squared fit to six parameters, namely the pixel coordinates of the centroid, length, width, and angle of the trail, and the peak value. We were then able to measure fluxes of trailed sources at each of the best-fit pixel coordinates of the centroids enclosed by a pill-shaped photometric aperture that consists of a rectangle having the average trail length and four times the average width as its length and width, respectively, and rotated by the average angle, and a semicircle on either side with its diameter the same as the width of the rectangle. The sky background was determined from an annulus surrounding the source, with the inner and outer limits 2$\times$ and 4$\times$ the aperture dimensions, respectively. We then derived the image zero-points in the SDSS photometric system using the Pan-STARRS 1 (PS1) catalogue \citep{2016arXiv161205560C} and the photometric transformation between the PS1 and SDSS systems by \citet{2012ApJ...750...99T}.

We show the general trend of the apparent magnitude of 323P in Figure \ref{fig:mapp}, in which the measurements are nightly mean values from the same filters and telescopes. Note that the data points from the ground telescopes and HST are a mix of different photometric apertures.

\subsection{Orbit Determination}
\label{ssec_orbit}

\begin{deluxetable*}{lr|c|c|c|c}
\tablecaption{Nongravitational Parameters of 323P/SOHO
\label{tab:NG}}
\tablewidth{0pt}
\tablehead{
\multicolumn{2}{c|}{Nongravitational Force Model} &
\multicolumn{3}{c|}{Nongravitational Parameters (au d$^{-2}$)} & RMS Residuals \\  \cline{3-5}
\multicolumn{2}{c|}{} & Radial $A_1$ & Transverse $A_2$ & Normal $A_3$ & (\arcsec)
}
\startdata
$\sim r_{\rm H}^{-n}$ & Slope index $n = 7.5$
                  & $\left(-4.56 \pm 0.53 \right) \times10^{-17}$
                  & $\left(-1.4271 \pm 0.0028 \right) \times10^{-18}$
                  & $\left(-1.25 \pm 1.54 \right) \times10^{-17}$
                  & 20.42 \\
 & 8.0
                  & $\left(-4.18 \pm 0.85 \right) \times10^{-18}$
                  & $\left(-3.0538 \pm 0.0046 \right) \times10^{-19}$
                  & $\left(-1.91 \pm 2.46 \right) \times10^{-18}$
                  & 12.05 \\
 & 8.5
                  & $\left(+2.11 \pm 1.58 \right) \times10^{-19}$
                  & $\left(-6.4991 \pm 0.0093 \right) \times10^{-20}$
                  & $\left(-2.73 \pm 4.55 \right) \times10^{-19}$
                  & 7.94 \\ 
 & 9.0
                  & $\left(+2.55 \pm 0.36 \right) \times10^{-19}$
                  & $\left(-1.3764 \pm 0.0021 \right) \times10^{-20}$
                  & $\left(-3.11 \pm 10.46 \right) \times10^{-20}$
                  & 12.22 \\ 
 & 9.5
                  & $\left(+9.55 \pm 0.90 \right) \times10^{-20}$
                  & $\left(-2.9019 \pm 0.0057 \right) \times10^{-21}$
                  & $\left(-1.97 \pm 26.54 \right) \times10^{-21}$
                  & 19.54 \\ 
\enddata
\tablecomments{
The same 247 out of 249 in total astrometric observations spanning an observed arc from 1999 December 12 to 2021 March 26 were included to obtain the nongravitational solutions for all of the models. We obtained the best fit with $n = 8.5$, as the O-C residuals of the solution show no obvious systematic trend and are within the $3\sigma$ of the measurement errors. See Section \ref{ssec_orbit} for detailed information.}
\end{deluxetable*}

We conducted astrometric measurements of 323P in our observations. The best-fit pixel coordinates of centroids of background stars (see Section \ref{ssec_phot}) were used to solve for plate constants of each image by least squares with the Gaia DR2 catalogue \citep{2018A&A...616A...1G}, whereby the astrometry of 323P and the associated uncertainty were obtained.

In addition to our astrometry, we received SOHO astrometry of the comet remeasured and provided by K. Battams, since prior to our observing campaign, SOHO was the only observatory that has been observing 323P around every perihelion passage of the comet since 1999. We then fed the astrometric data to the orbit determination code {\tt FindOrb} developed by B. Gray, which handles gravitational perturbation from the eight major planets, Pluto, the Moon, the most massive 16 main-belt asteroids, and relativistic corrections. The planetary and lunar ephemeris DE441 \citep{2021AJ....161..105P} was exploited. We weighted our astrometric measurements using the inverse square of the corresponding uncertainties, the worst of which does not exceed $\sim$0\farcs2. The SOHO astrometry was weighted using a scheme as a function of apparent magnitude of 323P (see details at \url{https://github.com/Bill-Gray/find_orb/blob/master/sigma.txt}).

At first, we attempted to fit a purely gravitational orbit to the astrometry of 323P. However, we found that only the observations from the nearest neighbouring apparitions could be fitted with the observed-minus-calculated (O-C) residuals within the measurement errors, and that the rest would have O-C residuals as large as $\sim$100$\sigma$ and exhibit an enormous systematic trend, resulting in a horrendous mean rms residual of 10\farcm2. We then proceeded to fit the orbit with inclusion of radial, transverse, and normal (RTN) nongravitational parameters, respectively denoted as $A_{1}$, $A_{2}$, and $A_{3}$, as additional free parameters in the orbit determination. The first nongravitational force model we tried was the one by \citet{1973AJ.....78..211M} assuming isothermal sublimation of water ice. Although there is a noticeable improvement in fitting the orbit, the majority of the astrometric data points from previous apparitions still have O-C residuals $\ga$50$\sigma$ with the systematic trend unsolved, and the mean rms residual of the fit is 2\farcm2. We thus conclude that the model by \citet{1973AJ.....78..211M} is inapplicable for 323P, suggestsing that the mass loss of the comet is highly unlikely related to free sublimation of water ice, or other typical cometary volatiles such as CO and CO$_{2}$, as they basically follow the inverse-square law in the observed heliocentric range of the orbit.

\begin{figure}
\epsscale{0.75}
\begin{center}
\plotone{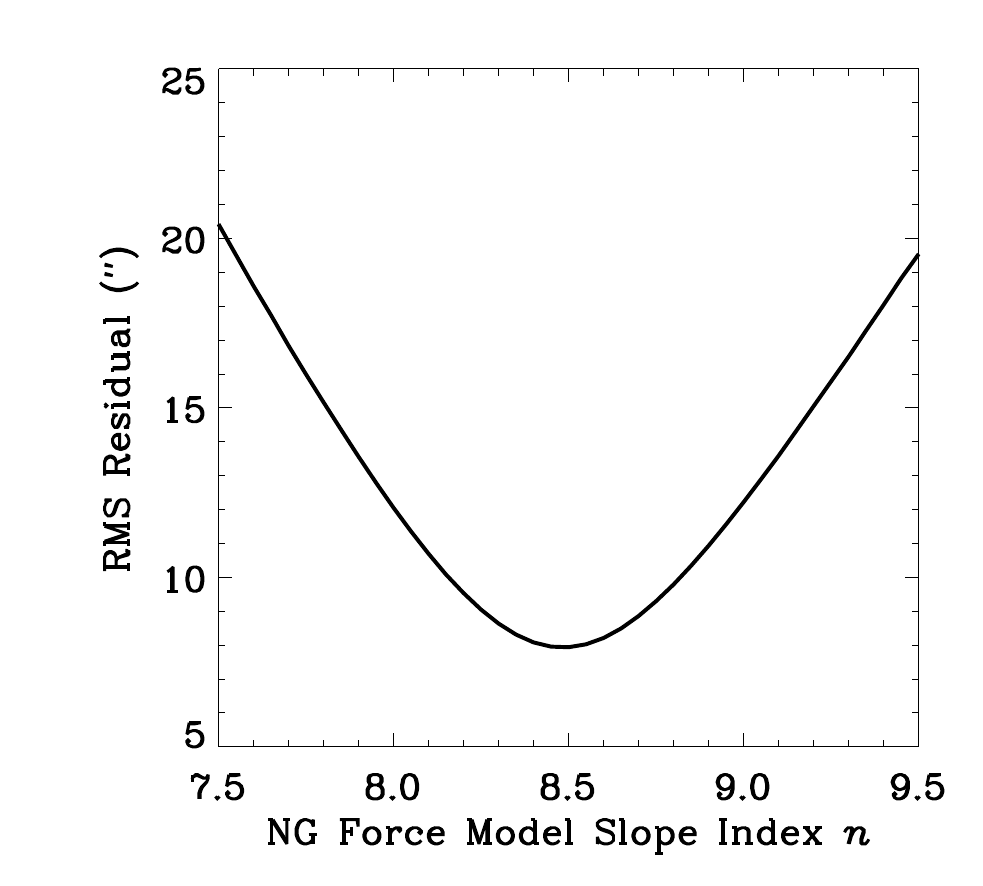}
\caption{
RMS residual of the best-fitted nongravitational orbital solution for 323P/SOHO versus the slope index of the power-law nongravitational force model $\sim r_{\rm H}^{-n}$. The local minimum of the rms residual is reached when $n = 8.50$.
\label{fig:NG_rms}
} 
\end{center} 
\end{figure}

Having recognised the shortcomings of the available nongravitational force models, we decided to adopt a nongravitational force model simply scaled as $r_{\rm H}^{-n}$, where $n$ is a constant power-law index. Following \citet{2020AJ....160...92H}, we searched for $n$ in a step size of 0.05 that would minimise the mean rms residual of the orbital solution. A rather satisfactory solution was found with $n = 8.50$, as the mean rms residual is merely 7\farcs9, and nearly all of the astrometric observations have residuals in agreement with the measurement uncertainty at the $3\sigma$ level (Figure \ref{fig:NG_rms} and Table \ref{tab:NG}). The only exception was two of the earliest SOHO observations from 1999, whose residuals are over ten times greater than those of the astrometry from the same apparition and therefore were rejected as outliers. The corresponding best-fitted RTN nongravitational parameters are $A_{1} = \left(+2.1 \pm 1.6 \right) \times 10^{-19}$ au d$^{-2}$, $A_{2} = \left(-6.499 \pm 0.009 \right) \times 10^{-20}$ au d$^{-2}$, and $A_{3} = \left(-2.7 \pm 4.5 \right) \times 10^{-19}$ au d$^{-2}$, where we can see that only the transverse component is statistically significant. The negative of the transverse nongravitational parameter, which does not necessarily imply whether the nucleus is in prograde or retrograde rotation \citep[][and citations therein]{1981AREPS...9..113S}, is nevertheless indicative of a secular acceleration, resulting in the orbital energy of the comet decreasing with time. 

\section{Discussion}
\label{sec_disc}

\subsection{Colour}
\label{ssec_clr}

\begin{figure}
\begin{center}
\gridline{\fig{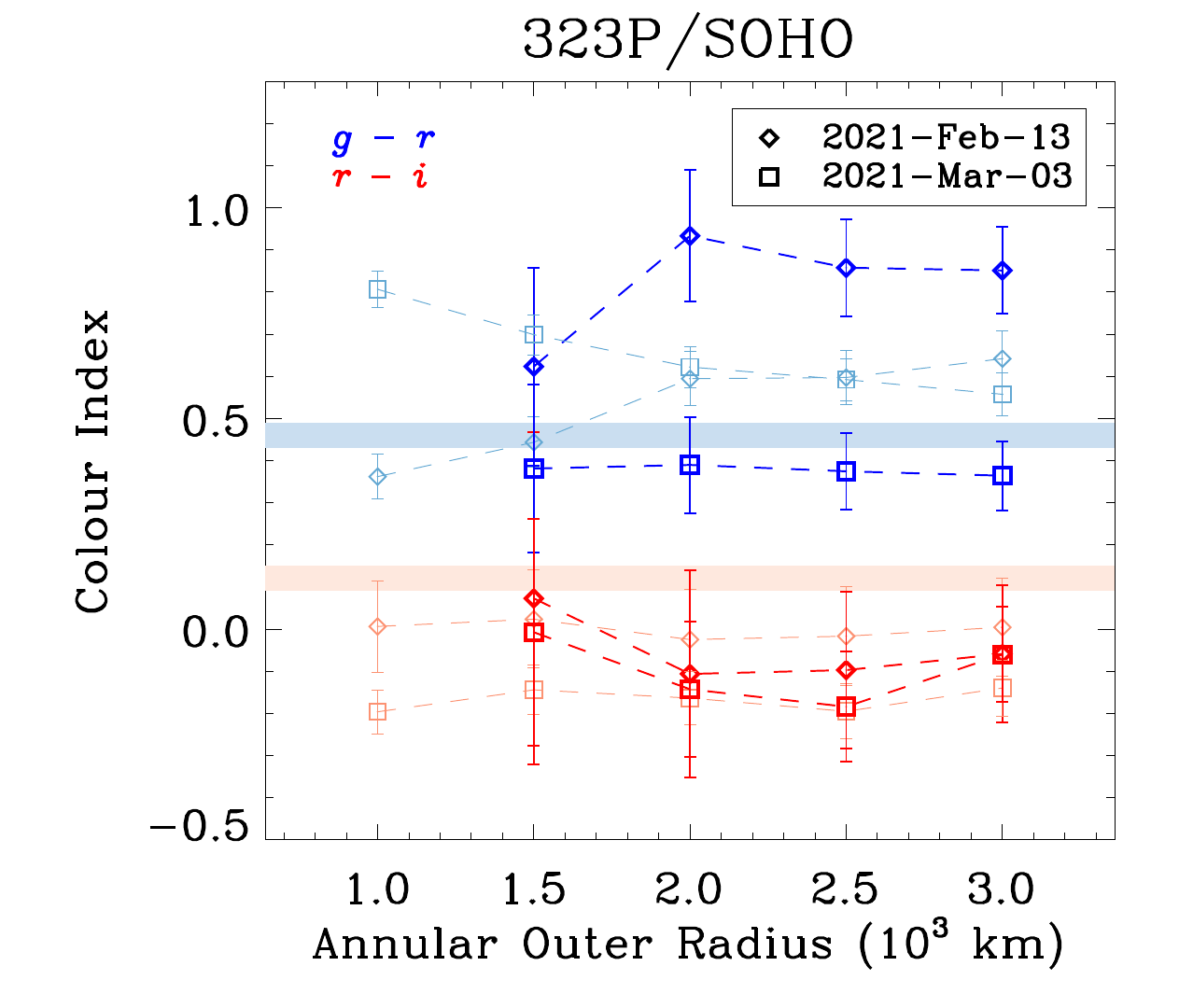}{0.5\textwidth}{(a)}
\fig{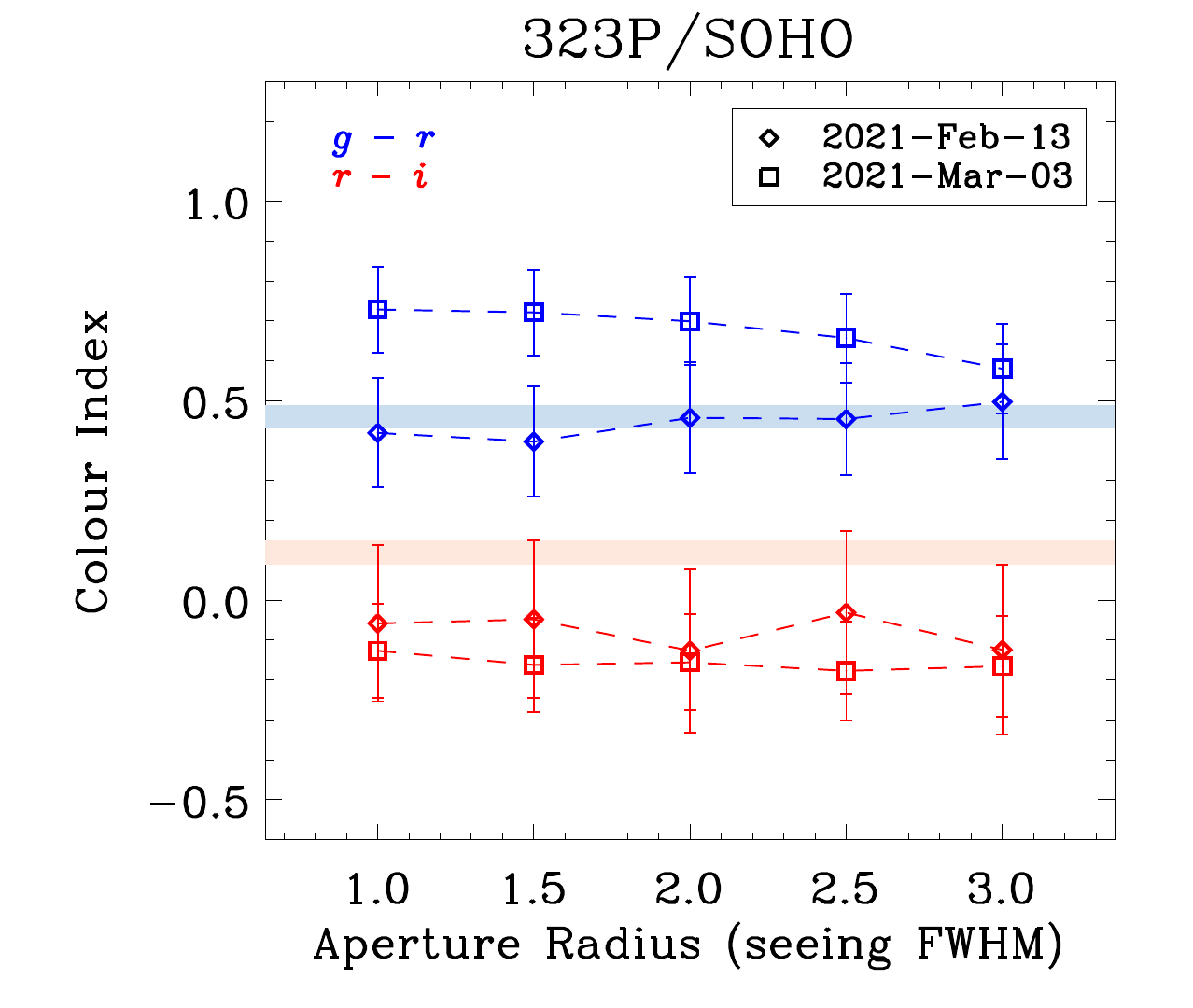}{0.5\textwidth}{(b)}}
\caption{
Colours of the dust ejecta (a) and the (b) nucleus of 323P/SOHO in the $g - r$ (blue) and $r - i$ (red) regimes. In the left panel, symbols in lighter colours are measurements from circular apertures of fixed linear radii, whereas those in darker colours are from annular apertures. In the right panel, the measurements with larger fixed seeing apertures are used as quality checks. Results from 2021 February 13 and March 3 are plotted as diamonds and squares, respectively. For comparison, the colour indices of the Sun are shown as the light blue and red stripes in the background (associated uncertainties included) in either panel.
\label{fig:color}
} 
\end{center}
\end{figure}

\begin{figure}
\epsscale{1}
\begin{center}
\plotone{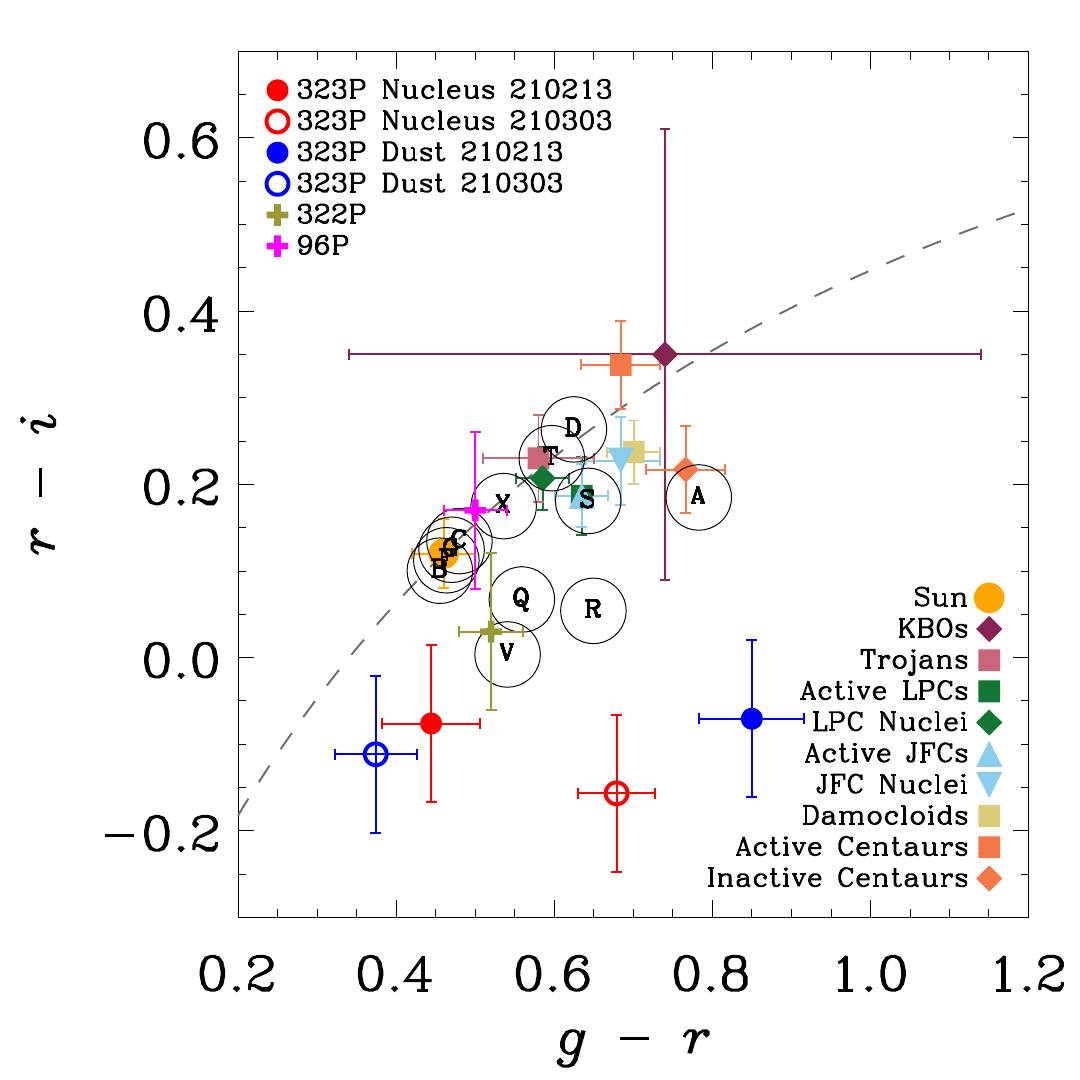}
\caption{
Colour comparison of the dust and the nucleus of 323P from the two GN epochs (2021 February 12 and March 3) with various solar system bodies, including the Sun \citep{2018ApJS..236...47W}, near-Sun comets 322P \citep{2016ApJ...823L...6K} and 96P \citep{2019AJ....157..186E}, Kuiper-belt objects \citep[KBOs;][and citations therein]{2012Icar..218..571S}, Centaurs, Jupiter Trojans, active Jupiter-family comets (JFCs) and their nuclei, active long-period comets and their nuclei, and Damocloids \citep[and citations therein]{2015AJ....150..201J}, in the $g - r$ vs $r - i$ space. Also plotted are typical colours of main-belt asteroid taxonomic classes \citep{2003Icar..163..363D}. Colour transformations to the SDSS system were performed according to \citet{2006A&A...460..339J}. The dashed curve is the locus of objects having linear reflectivity spectra. Objects below the curve have concave reflectivity spectra, otherwise convex, in the $g-i$ regime.
\label{fig:clr_comp}
} 
\end{center} 
\end{figure}

The Gemini multiband observations of 323P from 2021 February 13 and March 3 are the only dataset in our observing campaign that allowed for deriving the colour of the comet. In order to ensure that the obtained colour of the nucleus is free from any effect caused by the rotational modulation, we first calculated a preliminary folded lightcurve (see Section \ref{ssec_rot}) in which the colour indices of the nucleus were computed from the nightly mean values measured by the smallest fixed seeing aperture in the corresponding bandpasses. We then computed the predicted magnitude offsets caused by the rotational modulation and accordingly corrected the initial colour index values. The aforementioned steps were iterated once, after which the final corrections to the colour indices became less than their uncertainties, and there was no further improvement in the rms residual of the fit, and so there was no need to continue iterating the procedures. The colour corrections to the second Gemini observation were insignificant but rather noticeable to the first Gemini observation, because the originally planned imaging sequence from the latter was not completed due to a scheduling error. In addition, we computed the colour of the dust ejecta from the nightly mean values measured in a series of annuli with the inner radius fixed to be 1000 km from the nucleus projected at the distance of the comet, which meant to exclude the nucleus signal, and outer radii from 1500 to 3000 km in 500 km increments. We show the results in Figure \ref{fig:color}.

Intriguingly, the $g - r$ colour of the dust ejecta measured from the first Gemini observation appeared much redder than the Sun, but then became basically similar to the solar colour at the second GN epoch. As for the $r - i$ regime, the colour of the dust ejecta appeared to be bluer than that of the Sun at both epochs, with more certainty in the results from 2021 March 3. Because of the large uncertainties, we are uncertain about any spatial trends in the $g - r$ or $r - i$ colours of the dust ejecta from both of the epochs. We thus derived the weighted mean colour indices of the dust eject to be $g - r = 0.85 \pm 0.07$ and $r - i = -0.07 \pm 0.11$ on 2021 February 13, and $g - r = 0.37 \pm 0.05$ and $r - i = -0.11 \pm 0.07$ on March 3, in which the reported uncertainties are standard errors. We are unaware of other known comets having similar colours.

As for the colour of the nucleus, our measurements indicate that the way it varied is significantly different from the colour of the dust ejecta. On 2021 February 13, the nucleus appeared to have a solar-like colour in the $g - r$ regime, while the $r - i$ colour was possibly bluer than that of the Sun, yet the measured uncertainty is too large for a firm comparison. On March 3, however, in the $g - r$ regime the nucleus became significantly redder than the Sun, whereas the $r - i$ colour was measured to be noticeably bluer than that of the Sun. Given the uncertainties, the $r - i$ colour of the nucleus likely remained unaltered between the two GN epochs. For quality checks we measured additional photometry using larger fixed seeing apertures and applying corrections for the rotational modulation in exactly the same aforementioned way. The results were found to be consistent, although there appeared to exist a spatial trend due to the accumulative contamination from the dust ejecta around the nucleus as the aperture size increases. We computed weighted mean colour indices of the nucleus to be $g - r = 0.44 \pm 0.06$ and $r - i = -0.08 \pm 0.09$ on 2021 February 13, and $g - r = 0.68 \pm 0.05$ and $r - i = -0.16 \pm 0.05$ on March 3. 

To examine the reliability of the results on the colours of the nucleus and the dust ejecta, we remeasured the photometry of the comet in the Gemini data using different outlier rejection schemes and/or star catalogues, including the SDSS Data Release 12 \citep{2015ApJS..219...12A} and the ATLAS All-Sky Stellar Reference Catalog \citep{2018ApJ...867..105T}. However, the results did not alter beyond the noise level whatsoever, indicating that our derived colour of the comet should be trustworthy. Admittedly, a potentially important error that may be introduced to the colour of the nucleus is the correction for the rotational modulation. However, given the fact that the best-fit rotation period of the nucleus is robust and that data points from different bands line up smoothly within the uncertainties in the folded lightcurve (see Section \ref{ssec_rot} and Figure \ref{fig:nucrot}), we do not believe that our colour results of the nucleus are conspicuously biased.

In order to better understand the drastically different variations in the colours of the dust ejecta and the nucleus, we overplot measurements using circular apertures of fixed linear radii for the dust ejecta, which were obtained in exactly the same manner as for the results from annular apertures in Figure \ref{fig:color}a. We can see that the colours measured from the smallest two circular apertures are similar to what we obtained for the nucleus in Figure \ref{fig:color}b. As the size of the aperture increases, more signal from the dust ejecta was involved. On 2021 February 13, the $g - r$ colour turned redder as the aperture radius grows, suggesting that the dust ejecta was redder than the nucleus. However, on March 3, the trend became opposite, which means that the nucleus was bluer than the dust ejecta. Indeed, these inferences are in agreement with the measurements from annular apertures. As for the $r - i$ colour, we do not notice any strong spatial variation in the colour measurements using circular apertures, indicative of similar colours between the dust ejecta and the nucleus, which is also consistent with the annular aperture measurements.

Although several small solar system objects have been found to vary their colours due to mass-loss activity, whereby subterranean material is exposed as surface material is removed \citep[e.g.,][]{2019ApJ...882L...2M, 2020AJ....160...91H}, we are unable to find any other objects that resemble the temporal colour change in 323P. At the first GN epoch, the nucleus of 323P appeared to have a colour somewhat similar to the nucleus of near-Sun object 322P/SOHO \citep[$g - r = 0.52 \pm 0.04$ and $r - i = 0.03 \pm 0.06$;][]{2016ApJ...823L...6K}. Noteworthily, the nucleus of comet 96P/Machholz with its current perihelion distance $q = 0.12$ au was measured to have a fairly blue colour of $g - r = 0.50 \pm 0.04$ and  $r - i = 0.17 \pm 0.03$ by \citet{2019AJ....157..186E}. We conjecture that near-Sun objects are possibly bluer than many other small solar system bodies in general. In Figure \ref{fig:clr_comp}, we compare the $g -r$ and $r - i$ colours of 323P's dust and nucleus with the counterparts of various solar system bodies, whereby we can clearly notice the colour peculiarity of the object. In this regard, 323P appears to be unique in the overall population of the small solar system objects, and the physical mechanism responsible for its colour peculiarity is unclear. We suspect that the strange colour of 323P and the way it varied were possibly related to the mass loss around perihelion and intense solar heating in near-Sun environments. Anyway, we exhort future multiband observations of 323P and other near-Sun objects whenever opportunities ensue.

\subsection{Rotation}
\label{ssec_rot}

\begin{figure}
\epsscale{0.8}
\begin{center}
\plotone{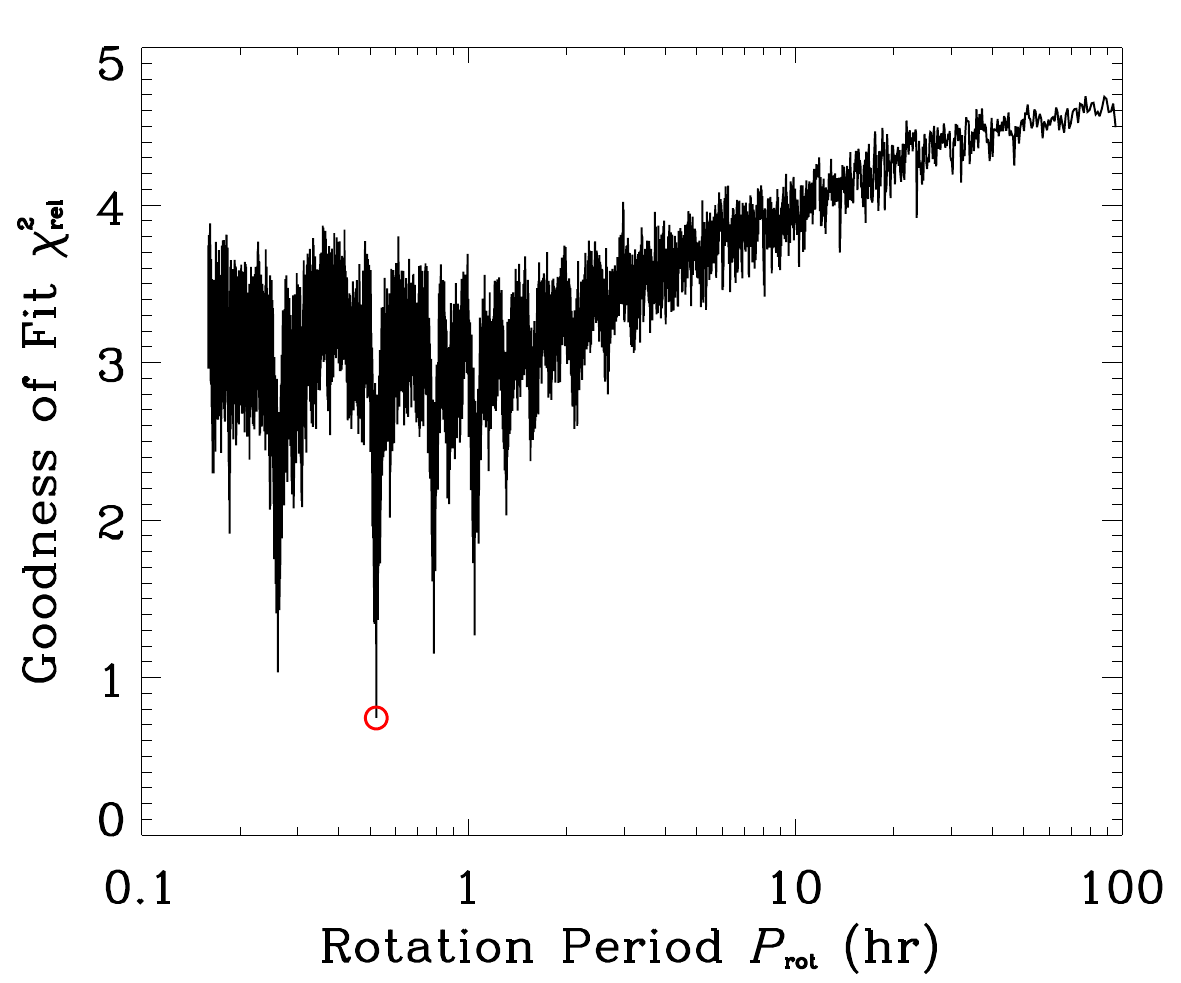}
\caption{
Periodogram of the nucleus of 323P/SOHO using relative lightcurves. The best solution (with a global minimum of $\chi_{\rm rel}^{2} \approx 0.7$, corresponding to a rotation period of $P_{\rm rot} \approx 0.522$ hr) is indicated by the red circle. The second best solution on the left side has half of the best period.
\label{fig:fnd_Prot}
} 
\end{center} 
\end{figure}

\begin{figure}
\begin{center}
\gridline{\fig{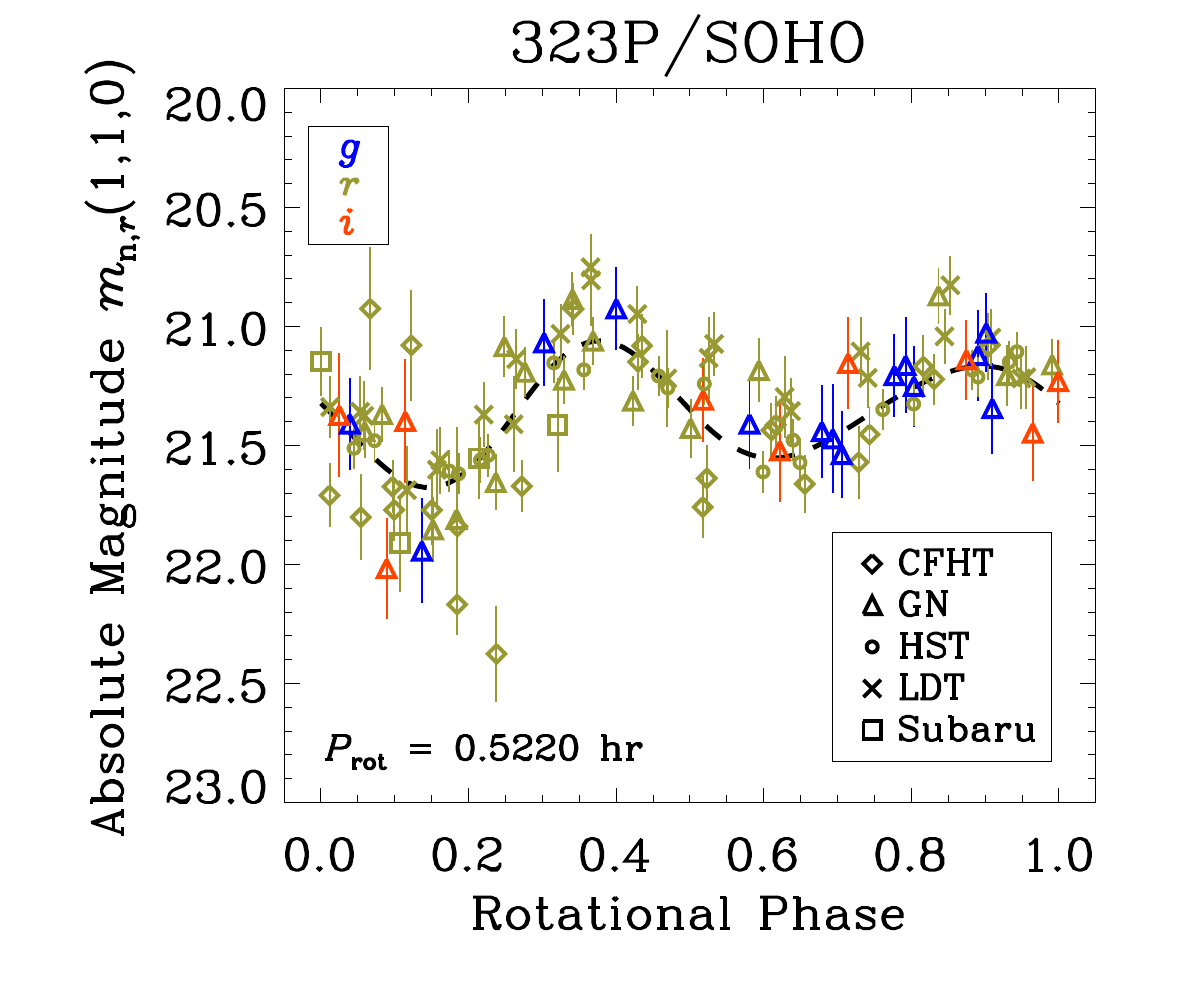}{0.5\textwidth}{(a)}
\fig{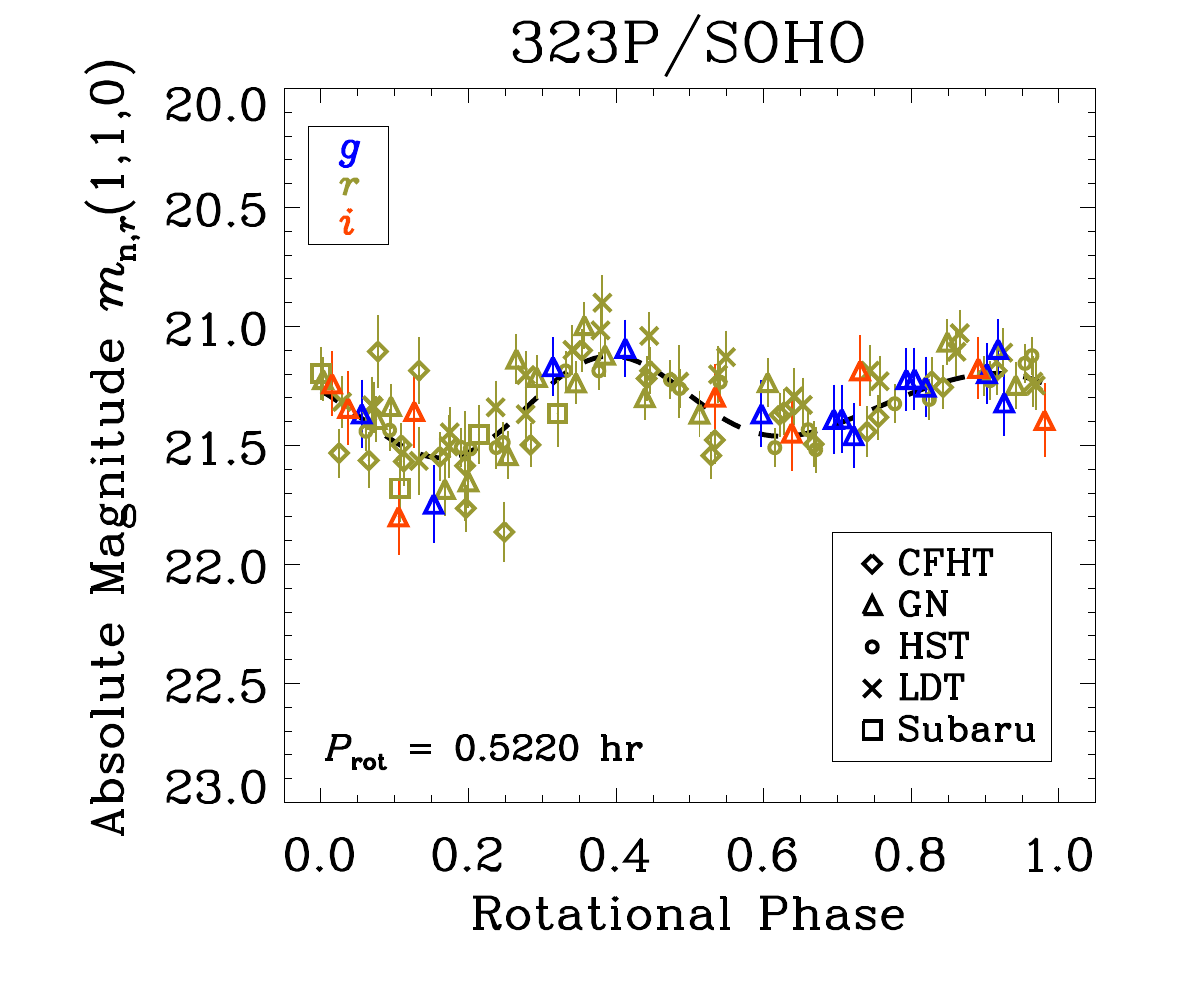}{0.5\textwidth}{(b)}}
\caption{
Folded rotational lightcurve of the nucleus of 323P/SOHO before (a) and after (b) removal of effects from the aspect change and the rotational lightcurve amplitude enhancement, with rotation period $P_{\rm rot} = 0.5220$ hr and the $1\sigma$ formal error thereof $\sim\!2 \times 10^{-6}$ hr. The epoch at which the earliest Subaru observation from 2020 December 21 was taken is set to be the reference. The black dashed curve is the best-fitted third order Fourier function described in Section \ref{ssec_rot}. Measurements from different observatories are distinguished according to the symbols shown in the legend. Note that the $g$ and $i$-band data in the plot (in colours different from the $r$-band data points) have already been converted to the $r$-band using the measured colour indices from the corresponding nights (see Section \ref{ssec_clr}).
\label{fig:nucrot}
} 
\end{center} 
\end{figure}

\begin{figure}
\epsscale{1.0}
\begin{center}
\plotone{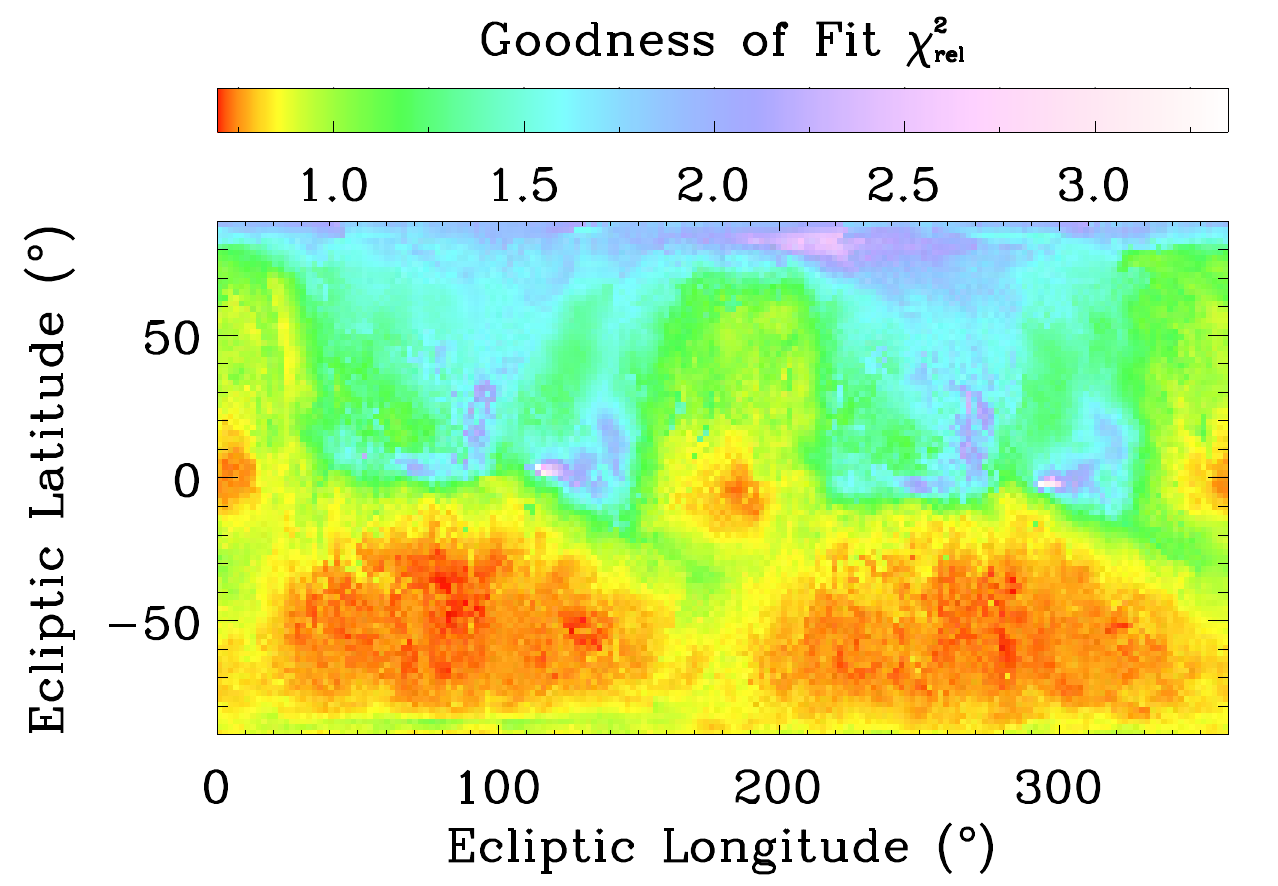}
\caption{
Goodness of fit $\chi_{\rm rel}^{2}$ as function of rotational pole orientation in terms of the J2000 ecliptic coordinates, obtained with the lightcurve inversion software package from DAMIT using the relative lightcurves. As indicated by the colour bar, the quality of the fit is colour coded. The result with the calibrated lightcurves is visually similar and is therefore omitted for brevity. Basically, the best pole solutions are concentrated around $\lambda_{\rm p} \approx 90\degr \pm 70$\degr~and $270\degr \pm 70$\degr~(possible mirror solutions) in ecliptic longitude, and $\beta_{\rm p} \approx -50\degr \pm 30$\degr~in ecliptic latitude.
\label{fig:pole}
} 
\end{center} 
\end{figure}

\begin{figure}
\epsscale{1.0}
\begin{center}
\plotone{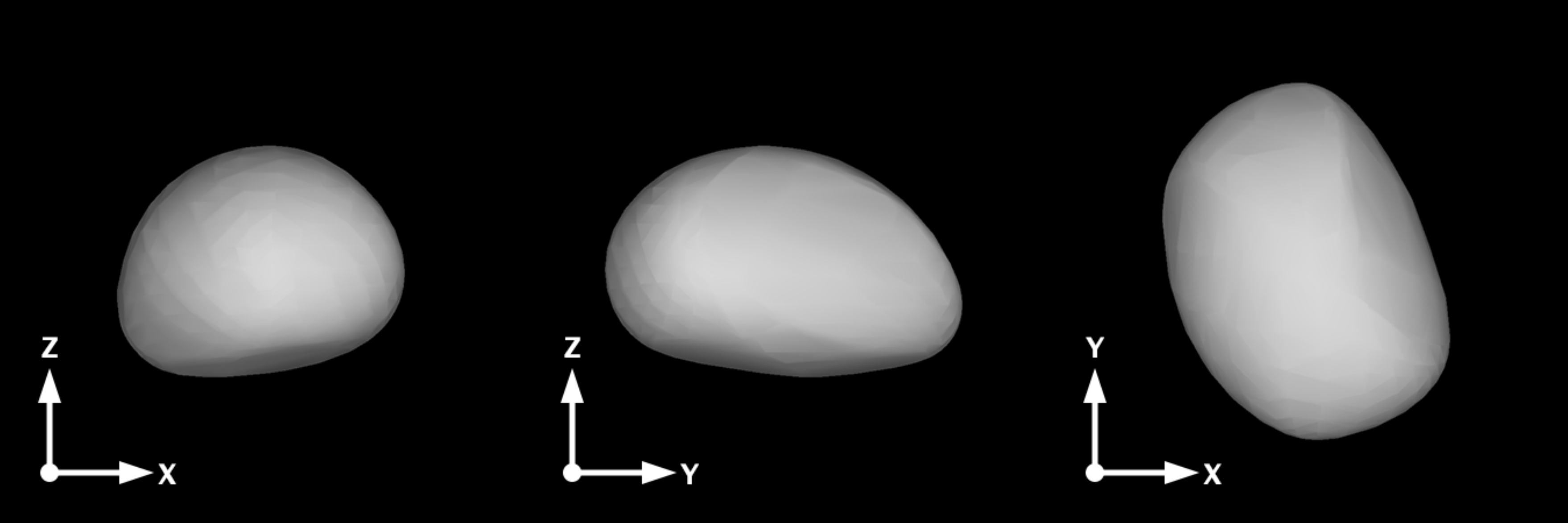}
\caption{
Convex shape model of the nucleus of 323P/SOHO in three orthogonal views, derived with the lightcurve inversion software package from DAMIT using the calibrated lightcurves. The left and middle panels show edge-on views of the nucleus that are 90\degr~apart from each other, with its spin axis (the $Z$-axis) pointing upwards, while the nucleus is viewed in a pole-on configuration (the line of sight along the negative $Z$ direction) in the right panel.
\label{fig:shape}
} 
\end{center} 
\end{figure}

While visually inspecting our obtained images of 323P, we noticed that the optocentre of the comet flickered in and out. Thus, we examined all of the photometric measurements with the smallest fixed seeing aperture from Section \ref{ssec_phot} to determine the nucleus rotational period. For the task, we adopted two different methods, the lightcurve inversion method \citep{2001Icar..153...24K,2001Icar..153...37K} and the Fourier analysis method \citep{1989Icar...77..171H}, and searched for the rotation period of the nucleus in a range of 0.25 to 150 rev d$^{-1}$. The observed epoch of each measurement was first converted to the Barycentric Dynamical Time (TDB) and then corrected for the light travel time. In the first method, we exploited a lightcurve inversion software package available from the Database of Asteroid Models from Inversion Techniques (DAMIT) project \citep{2010A&A...513A..46D}.\footnote{https://astro.troja.mff.cuni.cz/projects/damit/} A best-fit model to the lightcurves in relative brightness\footnote{In fact, the relative brightness was computed from our apparent magnitude measurements. Following \citet{2001Icar..153...24K} and \citet{2001Icar..153...37K}, we refer to them as relative lightcurves.} was computed for periods in the aforementioned interval, thereby obtaining the associated goodness of fit parameterised by relative chi-square $\chi_{\rm rel}^{2}$ \citep{2001Icar..153...24K}. We plot the search result in Figure \ref{fig:fnd_Prot}, in which we can see that the global minimum in $\chi_{\rm rel}^{2}$ is at a rotation period of $\sim$0.522 hr, and that the second best solution has the half best period. We repeated the same search for the lightcurves in calibrated brightness with and without measurements in the $g$ and $i$-bands converted to the $r$-band based upon the measured colour indices of the nucleus from the corresponding nights (see Section \ref{ssec_clr}). Consequently, we still found a global minimum in $\chi_{\rm rel}^{2}$ having a rotation period of $\sim$0.522 hr. 

We then switched to our code, which adopted the Fourier analysis method following \citet{1989Icar...77..171H} and has been repeatedly applied in analyses of super-fast rotators in particular \citep[e.g., ][]{2019ApJS..241....6C}. The distance-normalised lightcurve of the nucleus is expressed in the following form:
\begin{align}
\nonumber
m_{{\rm n}, r}\left[1,1,\alpha\left(t\right) \right] & \equiv m_{{\rm n}, r}\left[r_{\rm H}, {\it \Delta}, \alpha\left(t\right) \right] - 5 \log \left( r_{\rm H} {\it \Delta} \right) \\
& = H_{{\rm n}, r} - 2.5 \log {\it \Phi}_{\rm n} \left( \alpha \right) + \sum_{k=1}^{\mathcal{N}} \left[\mathcal{A}_{k} \sin \frac{2k\pi}{P_{\rm rot}} \left(t - t_{0} \right) + \mathcal{B}_{k} \cos \frac{2k\pi}{P_{\rm rot}} \left(t - t_{0} \right) \right]
\label{eq_phirot},
\end{align}
\noindent where $H_{{\rm n},r} = m_{{\rm n},r} \left(1,1,0\right)$ is the absolute $r$-band magnitude of the nucleus, including data points converted from other bandpasses, ${\it \Phi}_{\rm n}\left(\alpha\right)$ is the phase function of the nucleus normalised at zero phase angle, $t_{0}$ is some arbitrary referenced epoch, chosen to be the time of the earliest Subaru observation, $\mathcal{N} = 3$ is the degree of the Fourier series, and $\mathcal{A}_k$ and $\mathcal{B}_k$ ($k \in \mathbb{N}^{+}$ and $k \le \mathcal{N}$) are the Fourier coefficients. Four different phase functions -- the linear, $H,G$ \citep{1989aste.conf..524B}, $H,G_1,G_2$, and $H, G_{12}$ \citep{2010Icar..209..542M, 2016P&SS..123..117P} models were attempted and Equation (\ref{eq_phirot}) was solved numerically with the IDL-based Levenberg-Marquardt algorithm routine {\tt MPFIT} \citep{2009ASPC..411..251M}. However, we soon realised that the $H,G_1,G_2$ phase function model had to be discarded, because the slope parameters $G_1$ and $G_2$ could not be constrained. If we fixed either of the parameters, a solution with exactly the same minimum reduced chi-square ($\chi_{\nu}^2$) would be returned, suggesting that the quality of the measurements is not good enough to allow for solving phase models with more than a single slope parameters. Anyway, we refrain from discussion as to the phase function until in Section \ref{ssec_phi}. Regardless of which of the phase functions was adopted, we always found an obvious global minimum in the reduced chi-square at a rotation period of $\sim$0.522 hr, with a $1\sigma$ uncertainty of $\sim\!2 \times 10^{-6}$ hr. Using higher degrees of the Fourier series brought us no benefit, because the obtained $\chi_{\nu}^{2}$ did not decrease anymore with higher orders. As the best-fit rotation period is basically model-independent, we are therefore confident to conclude that the rotation period of the nucleus of 323P is $P_{\rm rot} \approx 0.522$ hr.

The rotational lightcurve of the nucleus of 323P phased with the best-fit period is plotted in Figure \ref{fig:nucrot}a, from which we can see that the $g$ and $i$-band data points produce a visually smooth folded lightcurve with the $r$-band data points within the corresponding errors. This suggests that the colours of the nucleus described in Section \ref{ssec_clr} are most likely unimpeachable. The folded rotational lightcurve clearly manifests double peaks and double troughs per period, with one trough slightly deeper than the other. The observed peak-to-trough amplitude, $\Delta m_{{\rm n},r} \approx 0.6$ mag, appeared to be somewhat scattered, possibly due to the drastic change in the viewing aspect (the angle between the spin axis orientation and the line of sight) and an amplitude-phase effect\citep[e.g.,][]{1990A&A...231..548Z,2019AJ....158..220L}, since both the ecliptic longitude and the phase angle of 323P varied by $\ga$70\degr~over the course of the observing campaign.

We investigated the spin axis orientation of the nucleus of 323P using the lightcurve inversion software package from DAMIT with the relative lightcurves. Initially, we treated the spin axis orientation expressed in the J2000 ecliptic coordinates as free parameters. However, we soon realised that the code would converge to distinct solutions with different initial guess values of the pole orientation. In order to understand how the quality of the fit varies with the pole orientation, we performed a raster scan of the entire $4\pi$ solid angle at a resolution of 2\degr~both in ecliptic longitude and latitude. The resulting goodness of fit $\chi_{\rm rel}^{2}$ for the relative lightcurves as a function of the spin axis oriented to a given direction in the J2000 ecliptic coordinates is shown in Figure \ref{fig:pole}. We can see that the best pole solutions are primarily concentrated around ecliptic longitude $\lambda_{\rm p} \approx 90\degr \pm 70$\degr~and ecliptic latitude $\beta_{\rm p} \approx -50\degr \pm 30$\degr. The region centred around a similar ecliptic latitude but at $\lambda_{\rm p} \approx 270\degr \pm 70$\degr~are possibly mirror solutions due to the 180\degr~spin longitude ambiguity commonly seen in the lightcurve inversion method \citep{2006InvPr..22..749K}. A narrower region around the vernal equinox of comparable goodness of fit also exists. We repeated the same aforementioned procedures for calibrated lightcurves, finding that the basic result is consistent. Given the orbit of 323P, we can conclude that the nucleus most likely spins in a retrograde manner.

Adopting the lightcurve inversion method, we derived a convex shape model of the nucleus of the comet for the spin axis oriented towards the aforementioned best region in the southern ecliptic hemisphere (Figure \ref{fig:shape}). The approximate axis ratios are $R_{2} / R_{1} \approx 0.8$ and $R_{3} / R_{1} \approx 0.7$. We tested with both the relative and calibrated lightcurves, finding that the resulting convex shape models of the nucleus are broadly unchanged, except that for the pole oriented at ecliptic longitude $\lambda_{\rm p} \la 80\degr$, the derived spin axis would become more aligned with the intermediate axis rather than the shortest one, which will place the nucleus rotation in an unstable regime. Given this, we prefer that the spin axis of the nucleus is oriented at some larger ecliptic longitude but nevertheless cannot rule out the possibility of it being in an unstable rotational state. Because of the great uncertainty, we opt not to overinterpret the results. The derived aspect ratio of the nucleus, $\sim$0.7, is relatively independent from the spin axis orientation, because this is primarily related to the rotational lightcurve amplitude, and the amplitude-phase relation is alike for various small solar system objects \citep[e.g.,][]{1990A&A...231..548Z,2019AJ....158..220L}. In comparison, Jupiter-family comets and near-Earth objects, both of which 323P belongs to, have aspect ratios $\sim$0.7 \citep{2004come.book..223L,2018A&A...611A..86C}, while main-belt asteroids have $\sim$0.8 \citep{2019AJ....157..164M}.

With the derived pole orientation and shape model, and based upon \citet[and citations therein]{1993Icar..106..563M} and \citet{1990A&A...231..548Z}, we applied a correction for effects from the aspect change and the rotational lightcurve amplitude enhancement by adding a coefficient to the Fourier series in Equation (\ref{eq_phirot}),
\begin{align}
\label{eq_amp}
\mathscr{A}\left({\it \Psi}, \alpha \right) & \equiv 
\frac{1 + \beta_{\rm A} \alpha}{2\log \left(R_2 / R_1 \right)} \log\left[\frac{\left(R_2 / R_3 \right)^2 \cos^2 {\it \Psi} + \left(R_2 / R_1 \right)^2 \sin^2 {\it \Psi}}{\left(R_2 / R_3 \right)^2 \cos^2 {\it \Psi} + \sin^2 {\it \Psi}} \right] \\
\nonumber
& = \left(1 + \beta_{\rm A} \alpha \right) \left\{ \frac{\sin^2 {\it \Psi}}{\sin^2 {\it \Psi} + \left(R_2 / R_3 \right)^2 \cos^2 {\it \Psi}} + \left(\frac{R_2}{R_1} - 1\right) \frac{\left(R_2 / R_3 \right)^2 \sin^2 {\it \Psi} \cos^2 {\it \Psi}}{\left[\sin^2 {\it \Psi} + \left(R_2 / R_3 \right)^2 \cos^2 {\it \Psi}\right]^2} + \mathcal{O}\left[\left(\frac{R_2}{R_1} - 1\right)^2 \right] \right\} \\
& \approx \left(1 + \beta_{\rm A} \alpha \right) \sin^2 {\it \Psi}
\label{eq_amp_approx},
\end{align}
\noindent which is practically the dimensionless rotational lightcurve amplitude normalised at aspect angle ${\it \Psi} = 90\degr$ and zero phase angle. Here, $\beta_{\rm A} \sim 10^{-2}$ deg$^{-1}$ is the slope of the rotational lightcurve amplitude enhancement with phase angle \citep{1990A&A...231..548Z}, and Equation (\ref{eq_amp_approx}) is an approximate form for Equation (\ref{eq_amp}) through Taylor expansion in case of the shape being not overly elongated. We adopted either of the forms, varied the pole orientation and the shape of the nucleus, as well as the value of $\beta_{\rm A}$ within their respective possible ranges, and refitted the distance-normalised lightcurve of the nucleus, finding that the best-fit rotation period and parameters in the adopted phase models all remain unchanged within the corresponding uncertainty levels and that the general shape of the folded rotational lightcurve is unaltered, but that the scatter therein is visibly improved (Figure \ref{fig:nucrot}b).

Finally, we remark that the best-fit rotation period, $P_{\rm rot} \approx 0.522$ hr, is the shortest for any known comets. In contrast, rotation periods of other known comets are at least a few hours \citep{2004come.book..223L,2004come.book..281S,2017MNRAS.471.2974K}, with the prior fastest being $\sim$2.8 hr \citep[near-Sun object 322P/SOHO; ][]{2016ApJ...823L...6K}. However, similar rotation periods have been previously reported for asteroids of comparable sizes \citep[e.g.,][]{2019ApJS..241....6C}. The rotational state is permissible within the spin-rate limit having nonzero internal cohesive strength \citep{2007Icar..187..500H}. To see this, we use order-of-magnitude calculation to estimate the critical internal cohesive strength that would be needed to marginally hold the material of the nucleus \citep[c.f.][]{2004come.book..281S}:
\begin{equation}
\sigma_{\rm c} = 2\pi \rho_{\rm n} R_{\rm n}^{2} \left(\frac{\pi}{P_{\rm rot}^2} - \frac{1}{3} \rho_{\rm n} \mathcal{G}\right)
\label{eq_cs}.
\end{equation}
\noindent Here, $\rho_{\rm n}$ and $R_{\rm n}$ are respectively the density and effective radius of the nucleus, and $\mathcal{G} = 6.67 \times 10^{-11}$ m$^3$ kg$^{-1}$ s$^{-2}$ is the universal gravitational constant. Substituion with the bulk density in a range of $\sim$0.4-3 g cm$^{-3}$, common for solar system bodies, and the nucleus radius of the comet (see Section \ref{ssec_xs}), Equation (\ref{eq_cs}) yields $\sim$10-100 Pa for the cohesive strength of the nucleus of 323P, which is consistent with previous studies on other cometary nuclei \citep[e.g.,][and citations therein]{2019SSRv..215...29G}.

\begin{figure}
\epsscale{1.0}
\begin{center}
\plotone{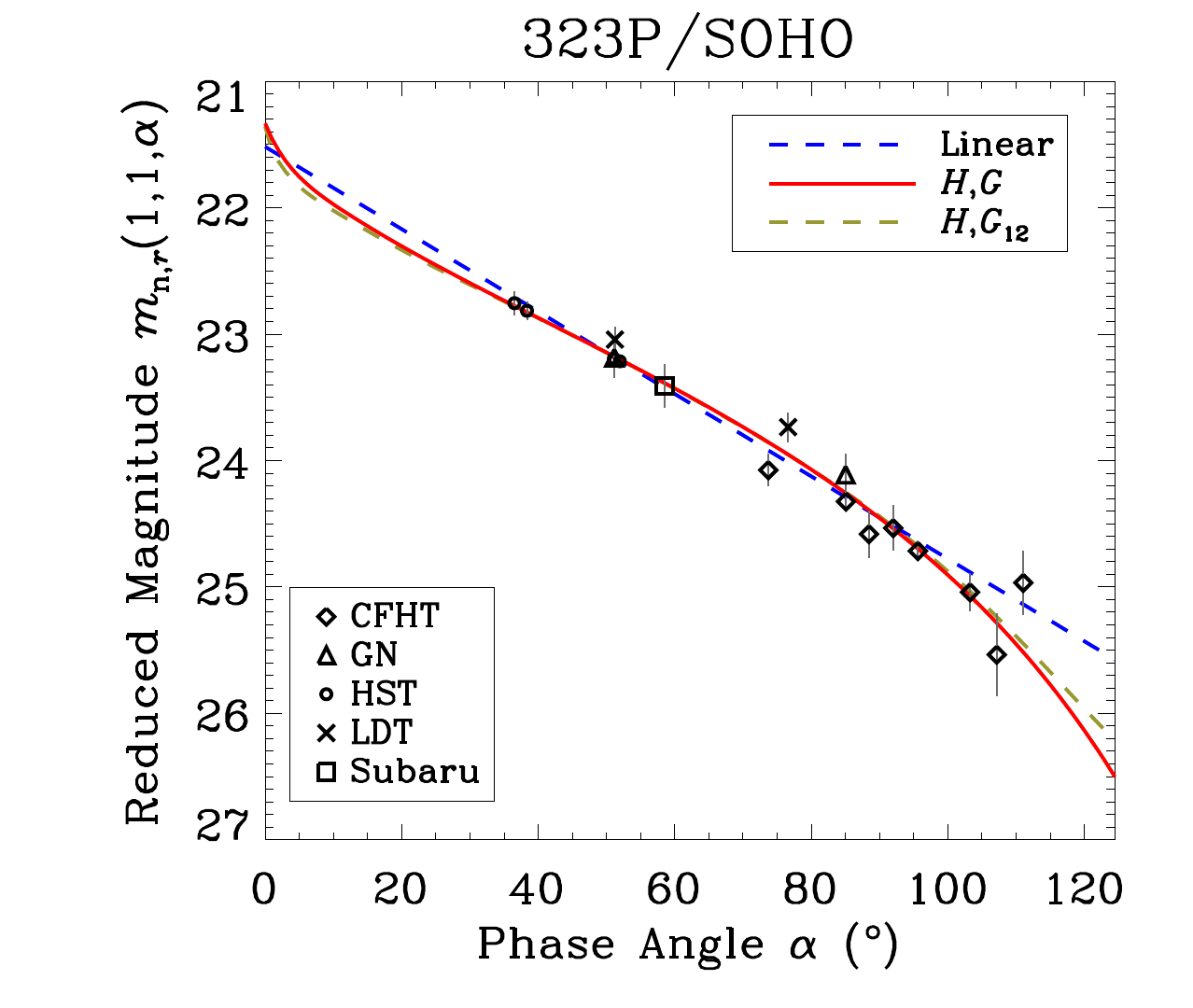}
\caption{
Phase function of the nucleus of 323P/SOHO. For clarity, the data points are weighted mean values of individual measurements from the same runs. As indicated in the legend, different symbols correspond to measurements from different telescopes. The best-fitted linear, $H,G$, and $H,G_{12}$ phase function models are shown as the blue dashed, red solid, and gold dashed lines, respectively. 
\label{fig:phi}
} 
\end{center} 
\end{figure}

\subsection{Phase Function}
\label{ssec_phi}

From Section \ref{ssec_rot} we also managed to obtain the best-fit linear, $H,G$, and $H,G_{12}$ phase models for the nucleus of 323P. The results are tabulated in Table \ref{tab:phirot} and shown in Figure \ref{fig:phi}, overplotted with weighted mean values of the photometric measurements from the same nights. The $H,G$ and $H,G_{12}$ models are our best solutions with the smallest $\chi_{\nu}^{2}$, thereby the smallest weighted rms residuals of the fit, despite that the former renders us a slightly better solution than the latter does (Table \ref{tab:phirot}).

Given the corresponding uncertainties, the best-fit slope parameters in the $H,G$ and $H,G_{12}$ phase models are both unexceptional in comparison to those of known small solar system bodies \citep[e.g.,][]{2012Icar..219..283O,2009Icar..202..134W}. In the case of the $H,G$ model, the slope parameter may serve as a proxy for taxonomic classification of small bodies \citep[e.g.,][]{2009Icar..202..134W}. However, the uncertainty is too large to meaningfully classify the taxonomy of 323P. Even if the uncertainty were sufficiently small, we argue that we still cannot taxonomically classify 323P solely based on its slope parameter because its peculiar colour is dissimilar to any of the known taxonomic complexes (see Figure \ref{fig:clr_comp}). We conjecture that there exists an ambiguity in taxonomic classification for the near-Sun population if one only relies upon slope parameters of their phase functions.

In spite of being worse than the other two models, we think that the best-fit linear phase function is still diagnostic to help draw comparisons with other small bodies. Our best-fit linear slope $\beta_{\alpha} = 0.0326 \pm 0.0004$ mag deg$^{-1}$ has no statistical difference from that of 322P \citep[$0.031 \pm 0.004$ mag deg$^{-1}$;][]{2016ApJ...823L...6K}, and is similar to nuclei of many other small solar system bodies such as Jupiter-family comets \citep{2017MNRAS.471.2974K} and near-Earth asteroids \citep{2013Icar..226..663H}. Besides, the geometric albedo is reported to be correlated with the slope of the linear phase function for comets and asteroids in the solar system \citep[e.g.,][]{2000Icar..147...94B,2013Icar..226..663H}. Judging from the relationship of near-Earth objects \citep{2013Icar..226..663H}, to which 323P belongs, we can estimate that the geometric albedo of the nucleus of 323P likely lies in a range of $\sim$0.1-0.2. 

\begin{deluxetable*}{c|c|c|c}
\tablecaption{Best-Fit Phase Functions of the Nucleus of 323P/SOHO
\label{tab:phirot}}
\tablewidth{0pt}
\tablehead{Model & 
\multicolumn{2}{c|}{Best-Fitted Parameters} & 
RMS Residuals (mag) \\ \cline{2-3}
 & Absolute $r$-band Magnitude $H_{{\rm n},r}$ (mag) & Slope Parameter\tablenotemark{$\dagger$} & }
\startdata
Linear & $21.52 \pm 0.02$ & $\beta_{\alpha} = 0.0326 \pm 0.0004$ &  0.1076 \\ \hline
$H,G$ & $21.33 \pm 0.08$ & $G = 0.17 \pm 0.04$ & 0.1010 \\ \hline
$H,G_{12}$ & $21.36 \pm 0.03$ & $G_{12} = 0.39 \pm 0.09 $ & 0.1014\\
\enddata
\tablenotetext{\dag}{The slope parameter of the linear phase model ($\beta_{\alpha}$) is in mag deg$^{-1}$, and the others are dimensionless.}
\tablecomments{All of the models give a best-fit rotation period of the nucleus $P_{\rm rot} = 0.5220$ hr, with $1\sigma$ formal uncertainty $\sim\!2 \times 10^{-6}$ hr, which is consistent with the standard deviation of the period solutions from different models. See Figure \ref{fig:phi} for comparison between the best-fitted phase function models and the data points. The associated uncertainties are $1\sigma$ formal errors derived from the covariance matrices of the best fits.}
\end{deluxetable*}

\subsection{Effective Scattering Cross-Section}
\label{ssec_xs}

\begin{figure}
\epsscale{1.0}
\begin{center}
\plotone{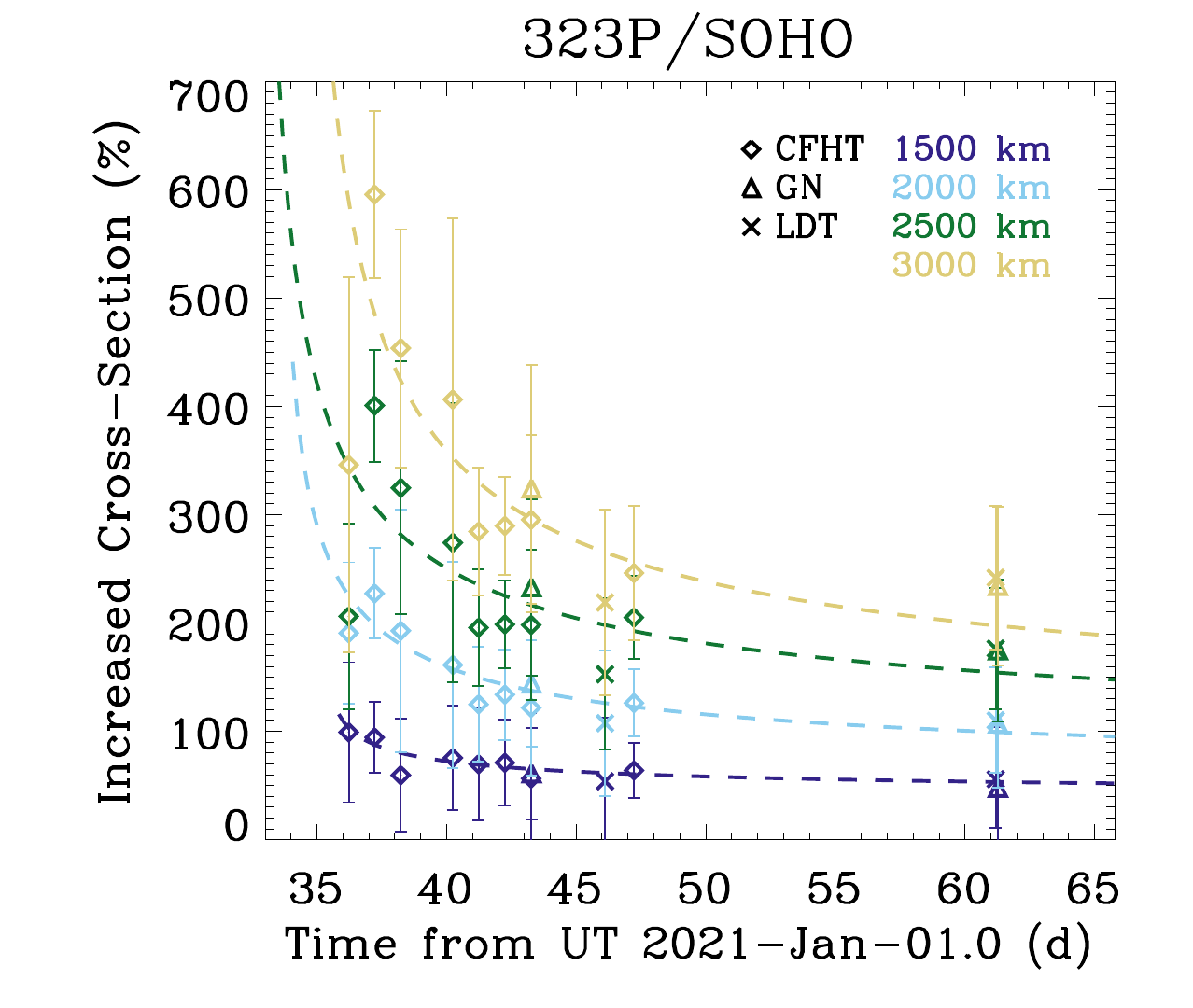}
\caption{
Effective scattering cross-section of the dust ejecta with respect to that of the bare nucleus as a function of time postperihelion, measured by a series of annular apertures whose inner radius was fixed to 1000 km from the nucleus projected at the distance of the comet, and the outer radii are as labelled in the plot. Only the $r$-band data points are used. Our best-fit physical models for the decline in the effective cross-section detailed in Section \ref{ssec_xs} are overplotted as dashed curves (colour coded corresponding to the annular size). Data points from different observatories are distinguished by different symbols.
\label{fig:eta}
} 
\end{center} 
\end{figure}

The effective scattering cross-section of the nucleus of 323P can be calculated from its absolute magnitude we obtained in Section \ref{ssec_rot} through
\begin{equation}
{\it \Xi}_{\rm n} = \frac{\pi r_{\oplus}^{2}}{p_{r}} 10^{0.4 \left(m_{\odot, r} - H_{{\rm n}, r} \right)}
\label{eq_xs_nuc},
\end{equation}
\noindent where $p_{r}$ is the geometric albedo of the nucleus, $m_{\odot, r} = -26.93 \pm 0.03$ is the apparent $r$-band magnitude of the Sun \citep{2018ApJS..236...47W}, and $r_{\oplus} = 1$ au is the mean Sun-Earth distance. Assuming $p_{r} = 0.15$ as appropriate for near-Sun objects \citep[e.g.,][]{2013AJ....145..133J,2019AJ....158...97M}, and also based on the relationship between the albedo and the slope of the linear phase function (see Section \ref{ssec_phi}), we obtain ${\it \Xi}_{\rm n} = \left(2.3 \pm 0.2 \right) \times 10^{-2}$ km$^{2}$, corresponding to a spherical nucleus radius of $R_{\rm n} = \left({\it \Xi}_{\rm n} / \pi \right)^{1/2} \approx 86 \pm 3$ m. Thus, the nucleus of 323P is likely slightly smaller than the one of 322P/SOHO, which is also a periodic near-Sun object having a nucleus radius of $\sim$125 m, if the same geometric albedo is assumed \citep{2016ApJ...823L...6K}. It is noteworthy that the reported uncertainties above do not incorporate the uncertainty in the geometric albedo, which is unconstrained from our observations. Thus, the actual uncertainty in the nucleus size will be most assuredly greater than what is given here, yet by an unknown amount. If the geometric albedo of the nucleus of 323P is twice larger than the assumed value here for instance, the effective radius will be reduced by a factor of $\sqrt{2}$.

The brightness excess of 323P in the postperihelion ground-based observations suggests additional contribution to the overall effective scattering cross-section from the dust ejecta. To get rid of the nucleus signal as much as possible, we picked photometric measurements from the annular apertures whose inner radius was set to 1000 km from the nucleus projected at the distance of the comet and out radii are between 1500 and 3000 km. The corresponding flux ratios between the dust ejecta in each of the annuli and the nucleus can be then calculated from
\begin{equation}
\eta = \frac{r_{\rm H}^{2} {\it \Delta}^{2}}{r_{\oplus}^{4} {\it \Phi}_{\rm n} \left( \alpha \right)} 10^{0.4 \left[H_{{\rm n}, r} - m_{r} \left(r_{\rm H}, {\it \Delta}, \alpha \right) \right]}
\label{eq_eta},
\end{equation}
\noindent in which $m_{r}$ is the $r$-band magnitude of the dust ejecta in each annulus, and we assumed the same phase function for the dust ejecta. If we instead assumed phase functions typical for cometary dust \citep[e.g.,][]{2007ICQ....29...39M,2011AJ....141..177S}, the resulting flux ratio in each annulus would keep rising with time, implying enduring dust release from the nucleus, which disagrees with our observations (see Section \ref{ssec_morph}). We therefore conclude that the observed dust ejecta of 323P is unlikely to have a phase function similar to those for typical cometary dust. We plot the flux ratios in different annuli in Figure \ref{fig:eta}, where we can see that the postperihelion effective scattering cross-section of the dust ejecta overall declined with time. Our interpretation is that the decline resulted from smaller sized dust being swept off from the photometric apertures by the solar radiation pressure while the larger counterpart remains therein. 

In the following we adopt a physical model to help understand the observed decrease in the effective cross-section of the comet. For simplicity, we assume a power law distribution for the dust size, ${\rm d}N \left(\mathfrak{a}\right) = {\it \Gamma \mathfrak{a}^{-\gamma} {\rm d} \mathfrak{a}}$, where ${\rm d} N \left( \mathfrak{a} \right)$ is the number of ejecta with radii from $\mathfrak{a}$ to  $\mathfrak{a} + {\rm d} \mathfrak{a}$, $\gamma$ is the constant power-law index, and coefficient ${\it \Gamma}$ is also a constant as long as the examined region is held fixed. At epoch $t_{\rm o}$, the outburst of the comet peaks and stops releasing new dust ejecta. Ignoring projection, we express the distance travelled by ejecta approximately as
\begin{equation}
\ell \approx \frac{\beta \mathcal{G}M_{\odot}}{2 r_{\rm H}^{2}} \left(t - t_{\rm o} \right)^{2}
\label{eq_dist}.
\end{equation}
\noindent Here, $M_{\odot} = 2 \times 10^{30}$ kg is the mass of the Sun, and
\begin{equation}
\beta = \frac{3 Q_{\rm pr} L_{\odot}}{16 \pi \rho_{\rm d} \mathfrak{a} \mathcal{G}M_{\odot} c}
\label{eq_beta}
\end{equation}
\noindent is the ratio between the solar radiation pressure acceleration and the local solar gravitational acceleration, in which $\rho_{\rm d}$ is the bulk density of the ejecta, $Q_{\rm pr} \approx 1$ is the scattering efficiency for radiation pressure \citep{1979Icar...40....1B}, $L_{\odot} = 3.8 \times 10^{26}$ W is the solar luminosity, and $c = 3 \times 10^8$ m s$^{-1}$ is the speed of light. Given a fixed $\ell$, the smaller the ejecta size, the shorter the time will be needed to cover the distance, in consequence of solar radiation being more efficient. From Equations (\ref{eq_dist}) and (\ref{eq_beta}), we can derive the critical size of ejecta that can travel distance $\ell$ as
\begin{equation}
\mathfrak{a}_{\rm c} \left(\ell \right) = \frac{3 Q_{\rm pr} L_{\odot}}{32 \pi \rho_{\rm d} \ell c} \left[\frac{t - t_{\rm o}}{r_{\rm H} \left(t \right)} \right]^{2}
\label{eq_ac}.
\end{equation}
The effective scattering cross-section of the ejecta within the circular area having radius $\ell$ can be expressed by
\begin{align}
\nonumber
{\it \Xi}_{\rm d} & = \int\limits_{\mathfrak{a}_{\rm c}}^{\mathfrak{a}_{\max}} \pi \mathfrak{a}^{2} {\rm d} N \left(\mathfrak{a} \right) \\
& \approx \frac{\pi {\it \Gamma}}{\gamma - 3} \mathfrak{a}_{\rm c}^{3 - \gamma}
\label{eq_XS}
\end{align}
\noindent for $\gamma > 3$ and $\mathfrak{a}_{\max} \gg \mathfrak{a}_{\rm c}$. Therefore, the effective cross-section of the dust ejecta inside an annulus region having projected radii from $\ell_1$ to $\ell_2$ is
\begin{align}
\nonumber
\Delta {\it \Xi}_{\rm d} \left(\ell_{1}, \ell_{2} \right) & = {\it \Xi}_{\rm d} \left(\ell_{2} \right) - {\it \Xi}_{\rm d} \left(\ell_{1}\right) \\
& \approx \frac{\pi}{\gamma - 3} \left[{\it \Gamma}\left(\ell_2 \right) \mathfrak{a}_{\rm c}^{3 - \gamma}  \left(\ell_2 \right) - {\it \Gamma}\left(\ell_1 \right) \mathfrak{a}_{\rm c}^{3 - \gamma}  \left(\ell_1 \right)\right]
\label{eq_dXS}.
\end{align}
\noindent We note from Equation (\ref{eq_eta}) that $\eta$ is essentially the ratio between the effective cross-sections of the dust ejecta and the nucleus, if they share a common geometric albedo. Thus, combining Equations (\ref{eq_ac}) and (\ref{eq_XS}), we derive
\begin{align}
\nonumber
\eta \left( t \right) & \approx \frac{\pi}{\left(\gamma - 3 \right) {\it \Xi}_{\rm n}} \left(\frac{3 Q_{\rm pr} L_{\odot}}{32 \pi \rho_{\rm d} c} \right)^{3 - \gamma} \left[\frac{{\it \Gamma}\left(\ell_2\right)}{\ell_{2}^{3 - \gamma}}  - \frac{{\it \Gamma}\left(\ell_1\right)}{\ell_{1}^{3 - \gamma}} \right] \left[\frac{t - t_{\rm o}}{r_{\rm H} \left( t \right)} \right]^{6 - 2\gamma} \\
& = \mathscr{C} \left[\frac{t - t_{\rm o}}{r_{\rm H} \left( t \right)} \right]^{6 - 2\gamma}
\label{eta_t},
\end{align}
\noindent where
\begin{equation}
\mathscr{C} = \frac{\pi}{\left(\gamma - 3 \right) {\it \Xi}_{\rm n}} \left(\frac{3 Q_{\rm pr} L_{\odot}}{32 \pi \rho_{\rm d} c} \right)^{3 - \gamma} \left[\frac{{\it \Gamma}\left(\ell_2\right)}{\ell_{2}^{3 - \gamma}}  - \frac{{\it \Gamma}\left(\ell_1\right)}{\ell_{1}^{3 - \gamma}} \right]
\nonumber
\end{equation}
\noindent is only a constant of no interest. For each annular aperture, we performed a least-squared fit to the data points in Figure \ref{fig:eta} using Equation (\ref{eta_t}), thereby finding $\gamma = 3.2 \pm 0.2$ and $t_{\rm o} = $ UT 2021 February $4\pm 3$. The reported uncertainties were propagated from errors in the repeated individual measurements. Instead, if the photometric measurements from circular apertures were directly fitted (with the aforementioned equations modified accordingly when necessary), we found $\gamma = 3.3 \pm 0.1$ and $t_{\rm o} = $ UT 2021 February $3 \pm 3$, which are consistent with the results with the annular apertures.

We were concerned about two major drawbacks in our model, which are the omission of the changing viewing geometry and the assumption of uniformly accelerated motions for the ejecta. To examine the reliability of the obtained $\gamma$ and $t_{\rm o}$, we attempted with modified models where we either regarded $r_{\rm H}$ as a constant or set the travelled distance equal to $\ell / \sin \alpha$ in Equation (\ref{eq_dist}), only to find that the best-fit results are not altered beyond the uncertainty levels. Therefore, we think that the results from our model, albeit simplistic, are trustworthy. Applying a more sophisticated model is beyond the scope of this paper. 

Nevertheless, the obtained $\gamma$ for 323P is identical to $\gamma = 3.2$ for SOHO-observed Kreutz sungrazing comets \citep{2010AJ....139..926K}, and is statistically indistinguishable from $\gamma = 3.2 \pm 0.1$ determined for disintegrated long-period near-Sun comet C/2015 D1 (SOHO) \citep{2015ApJ...813...73H}, $\gamma = 3.6 \pm 0.6$ for fragmenting comet 332P/Ikeya-Murakami \citep{2016ApJ...829L...8J}, and $\gamma = 3.3 \pm 0.2$ for disrupted active asteroid 354P/LINEAR \citep{2010Natur.467..817J}.

\subsection{Morphology}
\label{ssec_morph}

\begin{figure}
\epsscale{1.0}
\begin{center}
\plotone{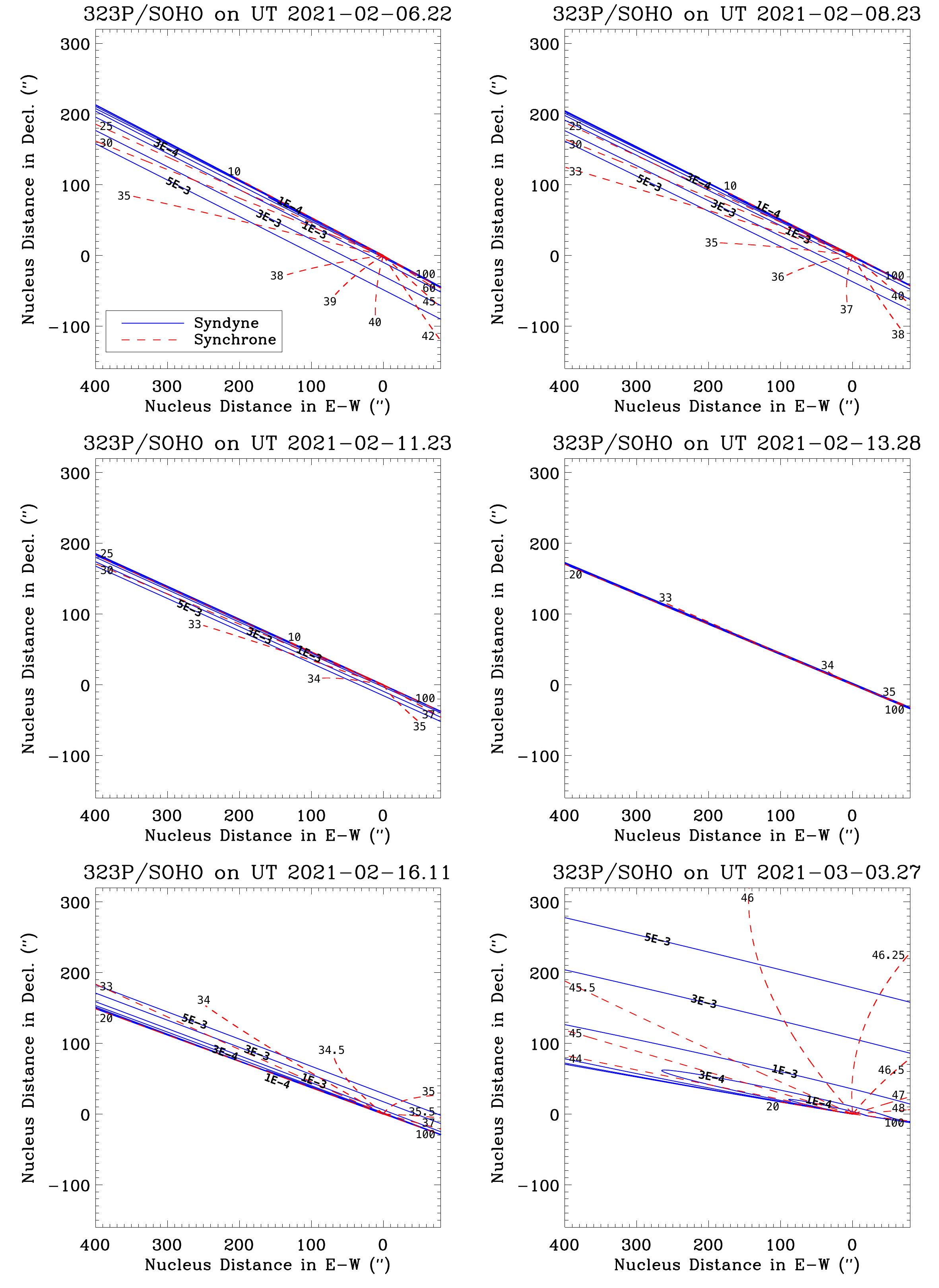}
\caption{
Finson-Probstein models of 323P corresponding to Figure \ref{fig:grnd_obs} for the ground-based observations (except the preperihelion Subaru observation in December 2020). As indicated in the legend, syndynes and synchrones are plotted as blue solid and red dashed curves, respectively. Syndynes of $\beta = 3 \times 10^{-4}$ and $10^{-4}$ (the most crowded two almost overlapped with each other) are left unmarked in the middle left panel for clarity. In the middle right panel, since Earth was closest to the orbital plane of the comet, all of the syndynes and synchrones almost collapsed to a single line, and thus we simply leave all of the syndynes therein unmarked for clarity.
\label{fig:FP_grnd}
} 
\end{center} 
\end{figure}

\begin{figure}
\epsscale{1.0}
\begin{center}
\plotone{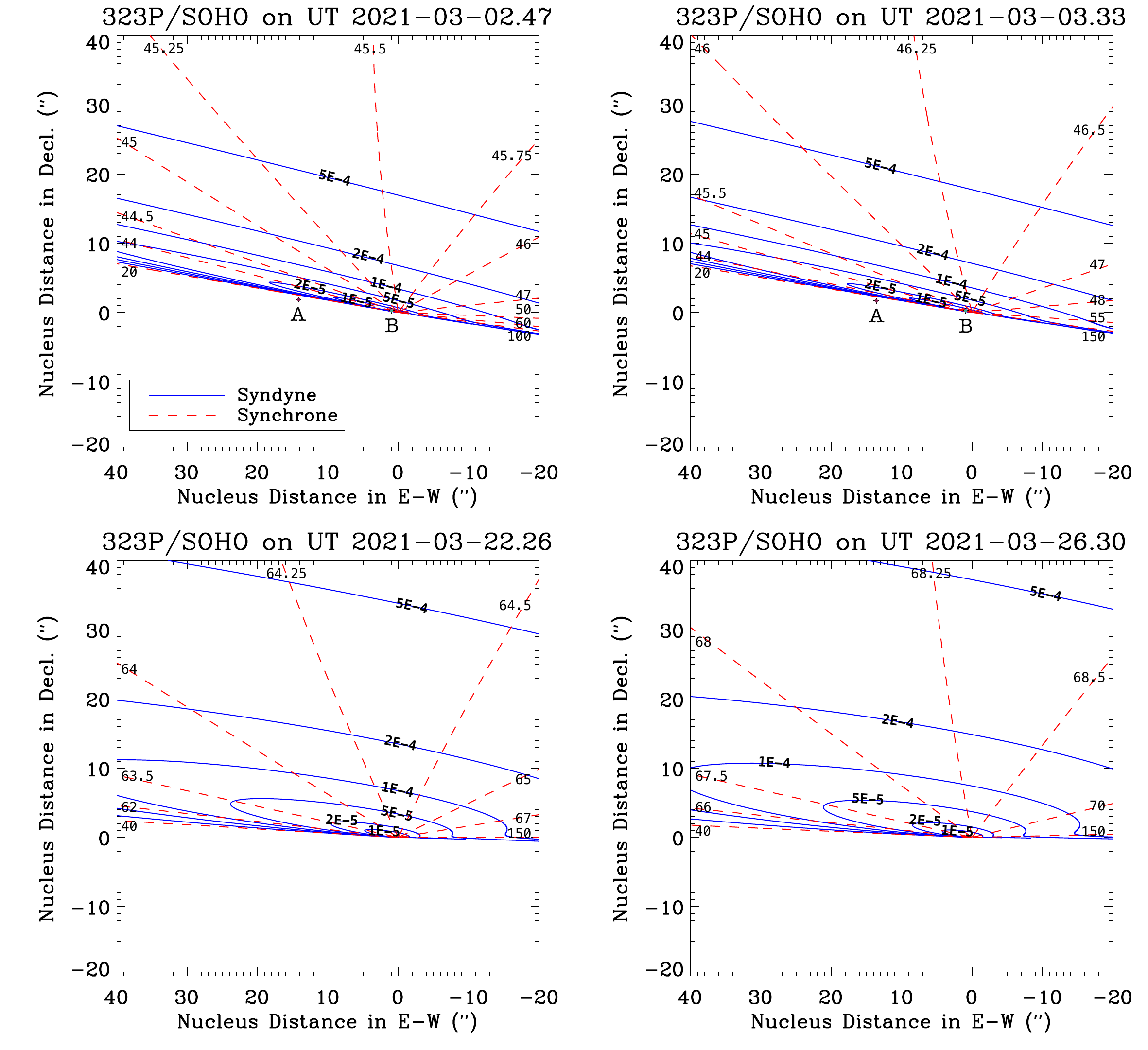}
\caption{
Finson-Probstein models of 323P corresponding to Figure \ref{fig:HST_obs} for the HST observations. As in Figure \ref{fig:FP_grnd}, syndynes and synchrones are respectively plotted as blue solid and red dashed lines. Also marked in the upper two panels are the relative positions of Fragments A and B from the primary along with their associated uncertainties.
\label{fig:FP_HST}
} 
\end{center} 
\end{figure}

We can infer physical properties of dust grains in the ejecta of 323P through morphologic analysis. The trajectory of a dust grain can be uniquely determined given the release time (here expressed in terms of time relative to the observed epoch), the $\beta$ parameter, and the initial ejection velocity $\bf{V}_{\rm ej}$ with respect to the cometary nucleus. 

The zeroth-order approximation to the dust morphology can be obtained through the syndyne-synchrone computation, in which syndynes are loci of dust grains subject to the same $\beta$ parameter but released from different epochs, synchrones are loci of dust grains released from the same epoch but subject to different values of the $\beta$ parameter, and the dust in the ejecta always leaves the nucleus with zero initial velocity \citep{1968ApJ...154..327F}. We plot the syndyne-synchrone diagrams at various covered epochs in our observing campaign (Figures \ref{fig:FP_grnd} \& \ref{fig:FP_HST}), to form comparison to the observations. What we found is that the observed ejecta of 323P is visually highly similar to a synchrone line with $\beta \la 2 \times 10^{-4}$ rather than a syndyne line, meaning that the observed dust ejecta was produced during a single episodic mass-shedding event within a day after the perihelion passage at a heliocentric distance of $r_{\rm H} = 0.04$ au. Compared to the outburst epoch estimated in Section \ref{ssec_xs}, the one derived by the syndyne-synchrone computation is far more accurate. Assuming a nominal bulk density of $\rho_{\rm d} = 1$ g cm$^{-3}$, we estimate from Equation (\ref{eq_beta}) that the ejecta of 323P mainly consists of dust grains having radii $\ga$3 mm. We note that the dust grains of this size are orders of magnitude larger than those ejected from typical comets \citep[$\beta_{\max} \sim 10^{-2}$;][]{2004come.book..565F} and the debris of disintegrated near-Sun comet C/2015 D1 (SOHO) \citep{2015ApJ...813...73H}. Besides, we observed that the comet displayed an extremely faint but discernible antitail feature roughly opposite to the position angle of the main dust ejecta in data collected from 2021 February 12 to March 3 (see Figure \ref{fig:grnd_obs}). Based upon the syndyne-synchrone approximation, we found that the antitail consisted of dust of $\beta \la 10^{-5}$ (corresponding to decimetre-size or larger) shed from the nucleus no later than November or December 2020, implying that although the observed main ejecta of the comet was produced from a massive mass-shedding event soon postperihelion, some mass loss has occurred even before the perihelion passage in January 2021.

We are aware that in reality, however, dust grains in the main ejecta of the comet must have left the nucleus with nonzero speeds, because our data gathered around the plane-crossing time on 2021 February 13 clearly show that the width of the ejecta appeared wider than the seeing FWHM in the normal direction of the orbital plane. The syndyne-synchrone model is no longer a good approximation when the ejection speed is sufficiently large \citep[c.f.][]{2019AJ....157..162H}:
\begin{equation}
\left|{\bf V}_{\rm ej} \right| \ga \beta \sqrt{\frac{\mathcal{G} M_{\odot}}{2r_{\rm H}}}
\label{ineq_vej}.
\end{equation}
\noindent Substituting, we find $\left|{\bf V}_{\rm ej} \right| \ga 20$ m s$^{-1}$. Such great speeds have never been previously seen in ejection of dust grains of similar sizes at comets or active asteroids, except in the case of (596) Scheila, whose transient activity was caused by an impact event \citep[and citations there]{2015aste.book..221J,2018SSRv..214...64L}. Therefore, we posit that our conclusions from the syndyne-synchrone approximation most likely remain valid. Nonetheless, we plan to employ our Monte Carlo cometary dust ejection model adopted and implemented from the one by \citet{2008Icar..193...96I}, in which the initial ejection velocities of dust grains are no longer omitted. We plan to present our detailed study of the dust environment and the interaction between meteoroids released from 323P and Earth in a separate paper in the future.

\subsection{Orbital Evolution}
\label{ssec_orbevo}

We desired to explore the dynamical pathway through which 323P became a near-Sun object and its potential dynamical fate. The negative of the transverse nongravitational acceleration implies that the orbital energy of the comet is ominously decreasing, which shrinks the semimajor axis of its heliocentric orbit. To have a basic sense about the orbital drift due to the nongravitational effect, we estimate the drift rate using Gauss' form of Lagrange's planetary equation \citep[e.g.,][]{2017AJ....153...80H}
\begin{align}
\nonumber
\bar{\dot{a}} & = \frac{P A_{2} r_{\oplus}^{n}}{\pi^2 a} \int\limits_{0}^{\pi} r_{\rm H}^{1-n} \left(\theta \right) {\rm d} \theta
\\
& = \frac{P A_{2}}{\pi \left(1 - e^2\right)^{n-1}} \left(\frac{r_{\oplus}}{a} \right)^{n} \sum_{k = 0}^{+\infty} \binom{n-1}{2k} \frac{\left(2k-1\right)!!}{\left(2k\right)!!} e^{2k}
\label{eq_da},
\end{align}
\noindent in which $n = 8.5$ (see Section \ref{ssec_orbit}) and $\theta$ is the true anomaly, and we have applied the binomial theorem to obtain the corresponding binomial coefficients. The binomial series apparently converges since the eccentricity of 323P is $e < 1$. Substituting with the best-fit orbital elements and $A_2$ from Section \ref{ssec_orbit}, Equation (\ref{eq_da}) renders us the orbital drift rate $\bar{\dot{a}} \approx -3 \times 10^{-2}$ au kyr$^{-1}$ at the present. Therefore, 323P seems doomed by eventually falling into the Sun if the negative transverse nongravitational parameter is maintained.

\begin{figure}
\epsscale{1.0}
\begin{center}
\plotone{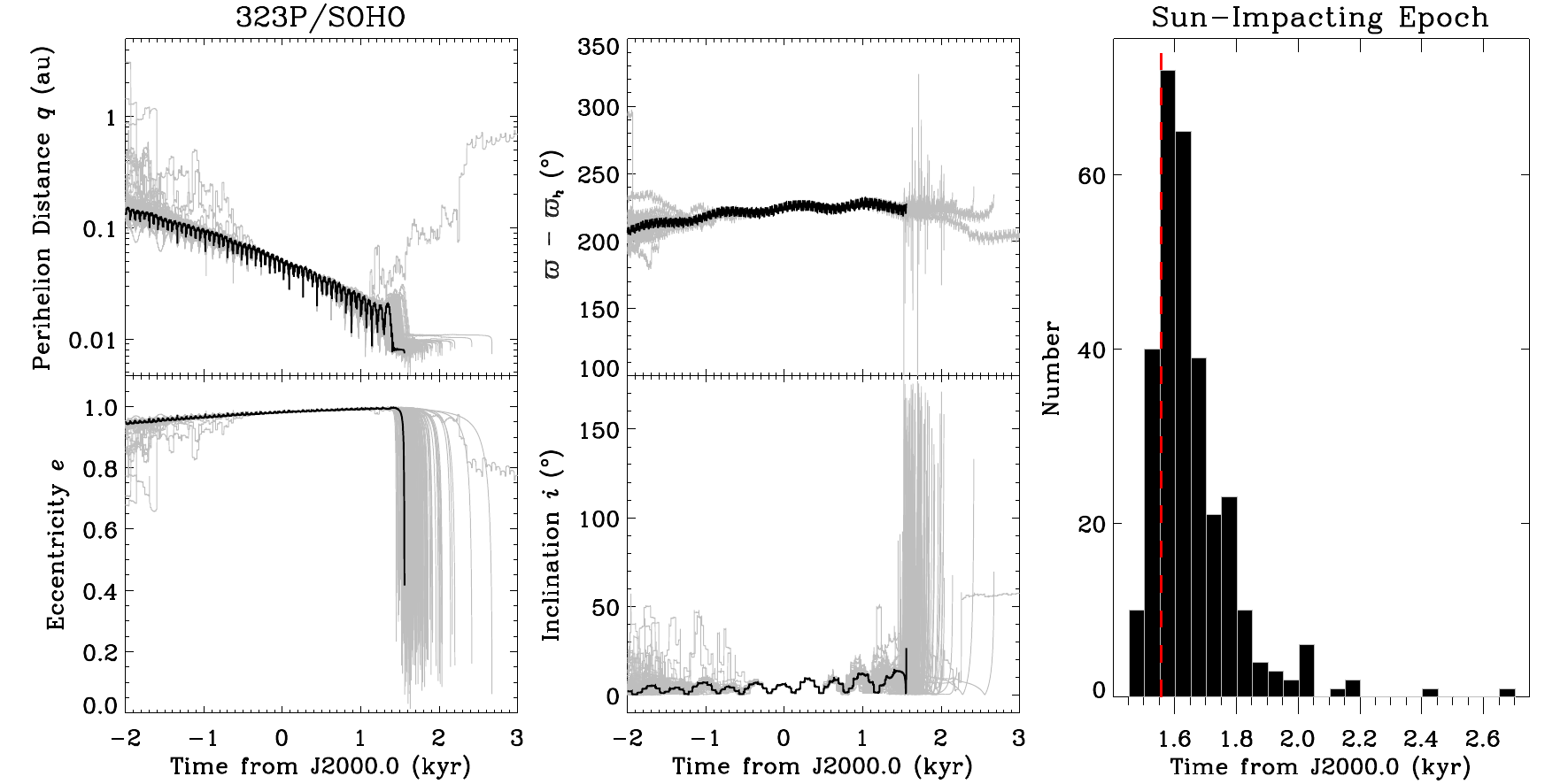}
\caption{
The left and middle panels show orbital evolution of the nominal orbit (dark) and the 300 MC variant clones (grey) of 323P/SOHO in terms of perihelion distance $q$, eccentricity $e$, inclination $i$, and the longitude of perihelion with respect to the one of Saturn, $\varpi - \varpi_{\saturn}$, from 2 kyr prior to J2000 to 3 kyr after, within which all of the clones but one collide with the Sun due to the $\nu_6$ secular resonance. We plot the distribution of the Sun-impacting epochs as the histogram on the right, where the vertical red dashed line represents the Sun-impacting epoch of the nominal orbit. From the dynamical statistics alone, we obtain that the comet has a likelihood of $\sim$99.7\% impacting the Sun $1.66 \pm 0.14$ kyr from J2000, in which the uncertainty is the standard deviation of the distribution.
\label{fig:orb_evo}
} 
\end{center} 
\end{figure}

We are fully aware that the orbit of 323P is highly chaotic because its orbit brings forth frequent close encounters with major planets. In order to incorporate the uncertainty in the orbit of 323P as much as we could, we exploited {\tt FindOrb} to add random noise to the astrometric measurements in accordance with the measurement uncertainties, followed by performing orbit determination iteratively for 300 times with completely the same nongravitational force model fitted as described in Section \ref{ssec_orbit}, whereby 300 Monte Carlo (MC) variant orbital clones were produced. We verified that the scatter in the orbital elements of these MC clones is comparable to the errors computed from the obtained covariance matrix of the nominal orbit. Note that our procedure may have underestimated the actual uncertainty in the nongravitational effect of 323P, because we did not incorporate potential secular variations therein but assumed constant nongravitational parameters for each of the orbital clones along with the nominal orbit. However, given that this factor cannot be constrained from the available astrometric observations, and the steep dependency upon the heliocentric distance, resulting in the nongravitational effect being negligible unless around perihelion passages, we thereby believe that errors introduced by our choice are unimportant in comparison to other uncertainties. 

Before proceeding to N-body integration of the nominal orbit and the 300 MC orbital clones, we calculated the Lyapunov timescale of 323P, denoted as $\tau_{\rm L}$, by means of the symplectic tangent map by \citet{1999CeMDA..74...59M}, where the distance between 323P and a nearby virtual particle was computed. In case of a chaotic system, the distance grows exponentially with time, and the Lyapunov timescale corresponds to the $e$-folding time. Our result is that the Lyapunov timescale of 323P is unsurprisingly short, only $\tau_{\rm L} \sim 50$-150 yr, which is similar to those of other near-Earth objects and Jupiter-family comets \citep{1998CeMDA..70..181T}. 

Bearing the obtained Lyapunov timescale in mind, we integrated backward and forward the nominal orbit along with the 300 MC orbital clones both for 5 kyr into the past and future separately utilising the numerical N-body integrator {\tt SOLEX12} \citep{1997CeMDA..66..293V}, in which perturbation from the eight planets, the Moon, Pluto, and the 16 most massive asteroids and post-Newtonian corrections were all taken into account, and the integration step size was adaptive, appropriate for handling close encounters.\footnote{The package is freely available at \url{http://www.solexorb.it/}.} For the forward integration, any collisions between the clones and the massive bodies in the force model were tracked by means of computing the closest approach distances. Whenever the distance was found to be no larger than the radius of some massive body in the force model, we treated it as an impact, with the impacting epoch recorded, and the corresponding clone would be removed from the subsequent forward integration. In the left panel of Figure \ref{fig:orb_evo}, we plot the orbital evolution of the clones of 323P in terms of perihelion distance, eccentricity, and inclination from 2 kyr before J2000 and 3 kyr after. 

First let us focus on the past orbital evolution of the comet. As expected, divergences in the perihelion distance, eccentricity, and inclination of the clones become noticeable starting from $\sim$0.5 kyr ago backward in time. Despite chaos in the orbit and the short Lyapunov timescale, over the past two millennia or thereabouts, all of the MC clones, including the nominal orbit of 323P, have been in prograde and highly eccentric heliocentric orbits with the overall trends of the perihelion distance decreasing over time. We verified that this behaviour is due to the fact that the comet has been in the $\nu_6$ secular resonance, which pumped up its eccentricity while the semimajor axis generally decreased with time (Figure \ref{fig:orb_evo}). Additionally, we calculated the Jupiter Tisserand parameter for each clone of the comet,
\begin{equation}
T_{\rm J} = \frac{a_{\jupiter}}{a} + 2\sqrt{\frac{a \left(1 - e^2 \right)}{a_{\jupiter}}} \cos i
\label{eq_TJ},
\end{equation}
\noindent where $a_{\jupiter}$ is the semimajor axis of Jupiter. What we found is that, over the past two millennia, all of the MC clones of 323P have been maintaining their Jupiter Tisserand parameters in a narrow range of $2 < T_{\rm J} < 3$. Altogether, while our backward integration cannot inform us about the source region of 323P (which should not be performed in this approach anyway, because otherwise integrating clones backwards for too long will lead to a manifest increase in entropy of the system backward in time, thereby violating the second law of thermodynamics), it is almost certain that, before becoming a near-Sun object, 323P used to have an unremarkable Jupiter-family orbit, whose perihelion distance has been generically decreasing over the past two millennia, superimposed with small-scale oscillatory patterns in a period of $\sim$60 yr primarily due to the near 5:2 mean-motion resonance between Jupiter and Saturn, which wobble the Sun around the barycentre of the solar system.

Next, we turn our attention to the dynamical fate of the comet. Our forward integration of the MC clones and the nominal orbit implies that the overall perihelion distance of the comet will highly likely continue shrinking, thereby turning the comet into a sungrazer. Except one of the MC orbital clone, all others including the nominal orbit (a likelihood of $\sim$99.7\%) eventually plunge into the Sun between $\sim$1.4 and 2.7 kyr from J2000 in our forward integration. Accordingly we calculated the mean Sun-impacting epoch from the 300 orbital clones to be $1.66 \pm 0.14$ kyr from J2000, where the reported uncertainty is the standard deviation. For these clones, the sudden changes in their orbital elements right before hitting the Sun (see Figure \ref{fig:orb_evo}) are caused by the adopted nongravitational force model.

One may wonder if the collision with the Sun of the comet would be strongly dependent upon the nongravitational effect. Had there been no other perturbing forces, the negative $A_2$ would tend to circularise the orbit of the comet \citep{2017AJ....153...80H}, which is opposite to the overall trend of the eccentricity shown in Figure \ref{fig:orb_evo}. Therefore, it is unlikely that the nongravitational effect leads to the collision with Sun. To have more certainty, we also integrated the orbit forward by completely neglecting its nongravitational acceleration, only to find that the dooming fate of the comet is unchanged in the next two millennia. This reinforces our conclusion that it is the $\nu_6$ secular resonance, rather than the nongravitational acceleration of the comet, as the primary dynamical mechanism decreasing the perihelion distance while increasing the eccentricity. This find is in line with \citet{1994Natur.371..314F} that the $\nu_6$ secular resonance is an important mechanism leading near-Earth objects to impact the Sun. 

Therefore, it is almost certain that 323P will cease to exist within the following two millennia due to the orbital evolution driven by the $\nu_6$ secular resonance. In reality, however, we expect that tidal disruption of the comet is bound to occur on its way falling into the Sun, even if it somehow manages to survive other fragmentation mechanisms. To see this, we calculate the Roche radius of the Sun for a fluid nucleus in synchronous rotation to be in a range of $\sim$2-4 $R_{\odot}$ with values of the bulk density between $\sim$0.4-3 g cm$^{-3}$ typical for solar system objects. From Figure \ref{fig:orb_evo}, we can estimate that the upper limit of the Roche limit will be reached in the next millennium or so, whereafter the nucleus of 323P will be likely torn apart attributed to the excessive tidal forces from the Sun.

\subsection{Fragments}
\label{ssec_frg}

\begin{figure}
\epsscale{1.0}
\begin{center}
\plotone{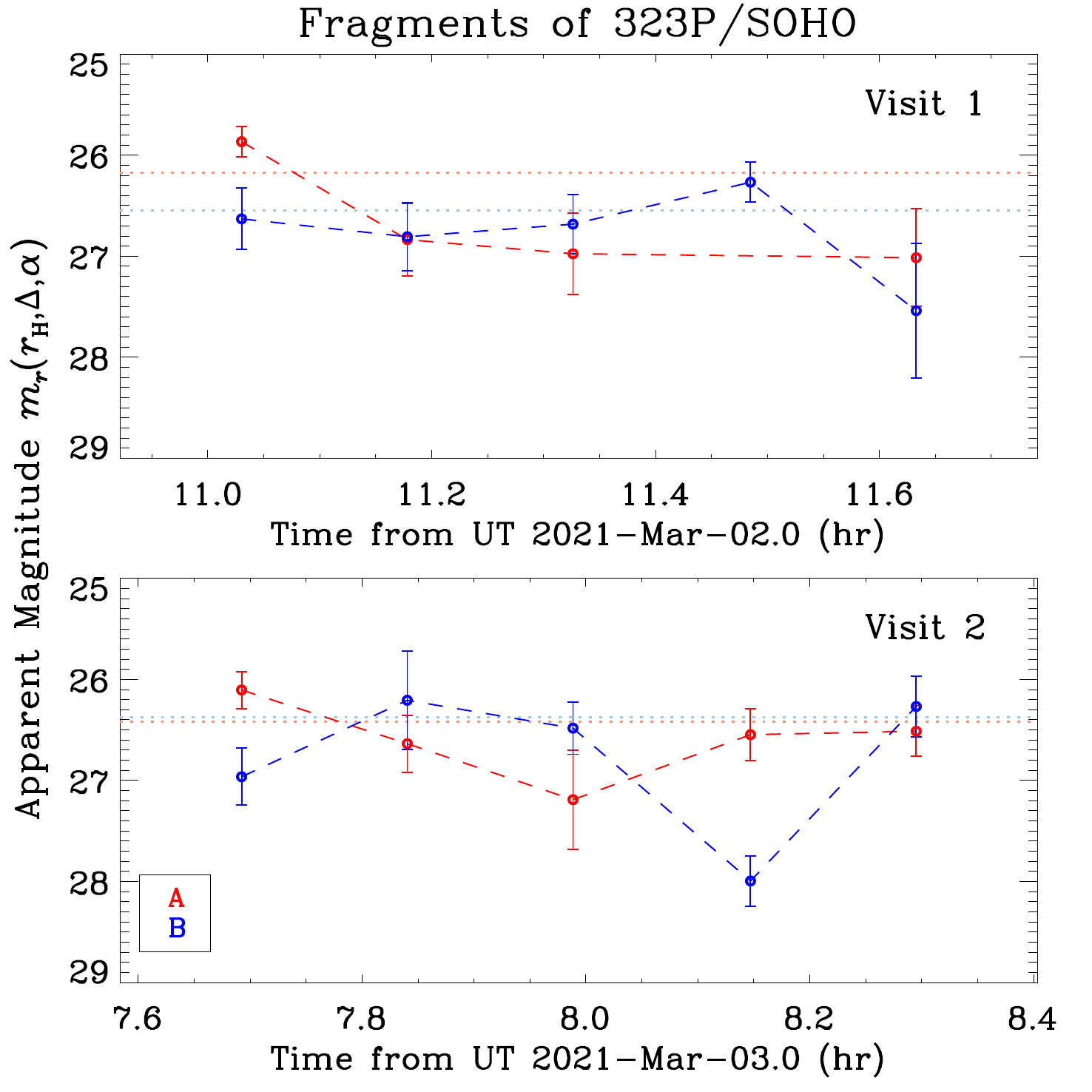}
\caption{
Apparent magnitude of Fragments A (red) and B (blue) of 323P from the first two HST visits on UT 2021 March 2 and 3, respectively. The horizontal dotted lines in the corresponding lighter colours are the weighted mean values of apparent magnitudes of the two fragments in either of the visits, whose errors are not shown in the plots for clarity.
\label{fig:lc_frg}
} 
\end{center} 
\end{figure}

\begin{deluxetable}{lc|c|c}
\tablecaption{Best-Fit Gravity-Only Orbital Solutions for Fragments A \& B of 323P/SOHO (Heliocentric Ecliptic J2000.0)
\label{tab:orb_frg}}
\tablewidth{0pt}
\tablehead{
\multicolumn{2}{c|}{Quantity}  & 
Fragment A  & 
Fragment B
}
\startdata
Perihelion distance (au) & $q$
       & $0.043 \pm 0.011$ & $0.057 \pm 0.018$ \\ 
Eccentricity & $e$
       & $0.9869 \pm 0.0049$ & $0.9793 \pm 0.0094$ \\ 
Inclination (\degr) & $i$
       & $5.61 \pm 0.31$ & $5.24 \pm 0.34$  \\ 
Longitude of ascending node (\degr) & ${\it \Omega}$
                 & $322.8 \pm 1.4$ & $324.6 \pm 1.8$ \\ 
Argument of perihelion (\degr) & $\omega$
                 & $355.32 \pm 0.78$ & $356.0 \pm 1.3$ \\ 
Time of perihelion (TDB)\tablenotemark{$\dagger$} & $t_\mathrm{p}$
                  & 2021 Jan $17.7 \pm 0.9$ 
                  & 2021 Jan $16.5 \pm 1.5$ \\
\hline
\multicolumn{2}{l|}{RMS residuals (\arcsec)}
& 0.049
& 0.060
\enddata
\tablenotetext{\dagger}{The corresponding uncertainties are in days.}
\tablecomments{Nine of the HST observations spanning an observing arc of 21.1 hr, from 2021 March 2 to 3, were used to obtain the orbital solutions, both of which are referred to a common osculation epoch of TDB 2021 March 3.0 = JD 2459276.5.}
\end{deluxetable}

\begin{figure}
\epsscale{1.0}
\begin{center}
\plotone{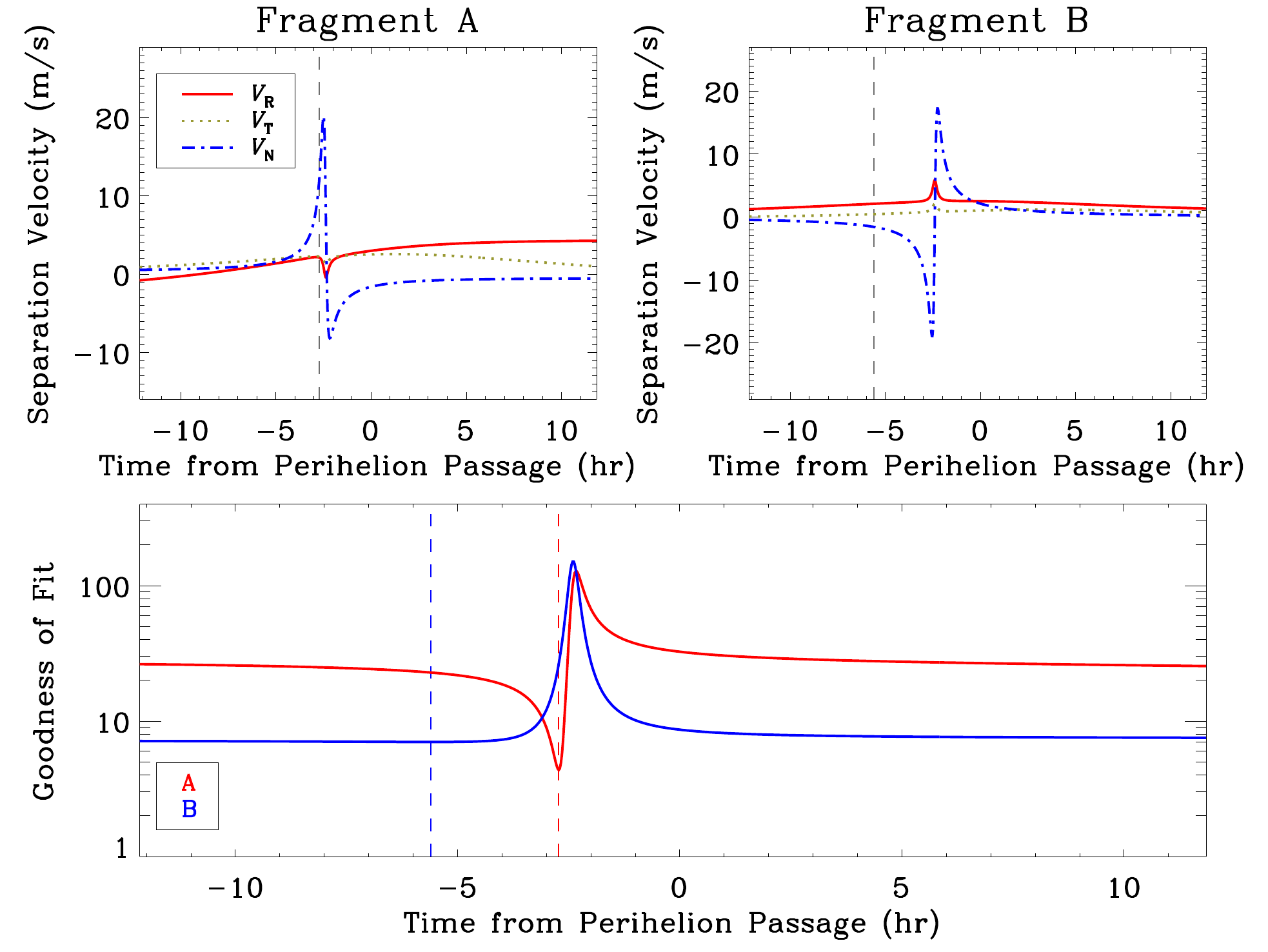}
\caption{
Best-fit RTN components of the separation velocities (upper two panels, discriminated by colours) and goodness of fit (bottom, also see Section \ref{ssec_frg}) of 323P's two fragments around the time of perihelion passage of the primary in 2021. For clarity the associated uncertainties in the separation velocity components are not shown. The minima in the goodness of fit for Fragments A and B (marked as the vertical dashed lines in the panels) are within $\sim$6 hr prior to the perihelion passage of the primary.
\label{fig:bestfit_frg}
} 
\end{center} 
\end{figure}

As we mentioned briefly in Section \ref{sec_res}, we identified two fragments of 323P in the HST/WFC3 images from 2021 March 2 and 3. First, we performed photometric measurements of the two fragments in the same way as we did for the primary in individual exposures from the first two HST visits. An image from the first visit in which Fragment A is stricken by a star trail was discarded. The brightness of the two fragments was then corrected for the aperture effect with the PSF model generated by {\tt TinyTim} \citep{2011SPIE.8127E..0JK}. We plot the measured apparent magnitudes of the two fragments in Figure \ref{fig:lc_frg}. We can see that although it is likely that both fragments exhibited brightness variations in the two HST visits, we did not see obvious repetitive patterns in their respective lightcurves due to insufficient coverage of the overall timespan. The mean apparent magnitude of Fragment A is $m_{{\rm A},r} = 26.17 \pm 0.13$ and $26.42 \pm 0.11$ on March 2 and 3, respectively, and for Fragment B we obtained $m_{{\rm B},r} = 26.55 \pm 0.13$ and $26.37 \pm 0.12$ in the two respective visits. The errors were propagated from individual measurement uncertainties. Assuming that the two fragments have a phase function in common with the one of their primary (Section \ref{ssec_rot}), we can estimate their absolute magnitude. Given the uncertainties, in which uncertainty from the phase function has been also included, we could not notice any obvious change in their intrinsic brightness between March 2 and 3. Thus, we computed mean values of their absolute magnitude during the two visits as $H_{{\rm A},r} = 24.32 \pm 0.09$ and $H_{{\rm B},r} = 24.44 \pm 0.09$ for Fragments A and B, respectively. Applying Equation (\ref{eq_xs_nuc}) and assuming the same geometric albedo as we used for the primary ($p_r = 0.15$), we estimate that the size of the two fragments are both $\sim$20 m in radius.

Assuming that the fragments remained intact throughout the HST observations, we gauge that their apparent magnitude would be $m_{r} \ga 27.5$ in the last two HST visits. By scaling the obtained S/N of the fragments accordingly from our photometric measurements, we immediately realise that both fragments would have S/N $\la 2$ in the individual exposures, which would place them below the detection threshold. The exact same conclusion can be reached if the WFC3 UVIS Imaging Exposure Time Calculator is exploited. Also bearing in mind the fact that the HST/WFC3 images are teemed with bombardments of cosmic rays, we therefore reckon that the loss of the two fragments in the last two HST visits is unsurprising attributed to their increasing faintness, albeit we cannot fully rule out the possibility that they have disintegrated and vanished in reality by the third HST visit.

Now we turn our focus onto the fragment kinematics. By performing conventional orbit determination, we confirmed that the two co-moving objects are fragments of 323P, in that they have heliocentric trajectories resembling the one of the primary. Thus, the possibility that the fragments had drifted beyond the FOV of the HST/WFC3 camera by the last two HST visits can be safely rejected, because their kinematics are not expected to be greatly different from those of the debris in the tail, and the parallactic displacement was diminishing, in consequence of the increasing observer-centric distance of 323P. We found no evidence that the fragments were subject to a nongravitational effect, in that pure-gravity solutions bring forth satisfactory O-C residuals without any systematic trend beyond the measurement errors (Table \ref{tab:orb_frg}). In fact, given the size of the fragments, we expect that their nongravitational parameters are $\la$10$^{-11}$ au d$^{-2}$ for a nongravitational acceleration arising from the solar radiation pressure. 

The rapid changing observing geometry of 323P in 2021 March prevented us from directly using the sky-plane component of the separation distances between the fragments and the primary nucleus as the proxy to study the fragment kinematics \citep[e.g.,][]{2016ApJ...829L...8J,2021AJ....162...70Y}, otherwise one would find that both of the fragments were spatially approaching the primary, resulting in a precarious argument that they were trapped in the gravitational field of the primary. To evaluate the likelihood, we estimate the apparent angle subtended by the Hill radius to the HST
\begin{equation}
\vartheta_{\rm H} = \frac{qR_{\rm n}}{{\it \Delta}} \left(\frac{4\pi \rho_{\rm n}}{9 M_{\odot}} \right)^{1/3}
\label{eq_theta_Hill},
\end{equation}
\noindent whereupon we obtain $\vartheta_{\rm H} \sim 1$ mas, utterly unresolvable in the HST/WFC3 images. We can therefore assuredly deduce that the two fragments cannot be gravitationally bound to the primary as they are way beyond the Hill sphere of the latter.

Rather, we applied a modified cometary fragmentation model that was initially devised by \citet{1977Icar...30..574S,1978Icar...33..173S}. With mutual gravitational perturbation between the fragments and the primary nucleus ignored, the trajectory of either of the fragments can be parameterised by the separation velocity with respect to the primary (${\bf V}_{\rm sep}$, expressed in terms of the RTN components $V_{\rm R}$, $V_{\rm T}$, and $V_{\rm N}$), and fragmentation epoch $t_{\rm frg}$. The best-fit split parameters were obtained with {\tt MPFIT} by fitting the apparent angular separation between the fragments and the primary nucleus decomposed in the J2000 equatorial east-west and decl. directions. The goodness of fit is given by
\begin{equation}
\chi_{\rm frg}^{2} \left(t_{\rm frg}, V_{\rm R}, V_{\rm T}, V_{\rm N} \right) = {\bm \xi}^{\rm T} {\bf W} {\bm \xi}
\label{eq_chi2_frg},
\end{equation}
\noindent where ${\bm \xi}$ is the corresponding O-C residual vector, and ${\bf W}$ is the weight matrix assigned based on the obtained astrometric observation errors. For uncorrelated observation errors, ${\bf W}$ is simply a diagonal matrix.

During the HST observations, neither of the fragments showed detectable apparent motion with respect to the primary, and therefore, in essence we only have their positions at two different epochs in the cometocentric coordinate system. Given the measurement errors, we realised that it was impractical to straightforwardly treat all of the split parameters as free parameters to be solved in our code. By comparing the apparent positions of the two fragments relative to the primary with the syndyne-synchrone grid, we immediately ascertained that the separation velocity must play an important role for Fragment A, because it was situated at a locus where no syndyne or synchrone lines could reach at all (Figure \ref{fig:FP_HST}). 

Through initial tests, we found that feeding different initial guess values to our code would lead to convergence to statistically different best-fit split parameters with comparable goodness of fit, in particular the split epoch, although the best-fit separation speed is always $\left|{\bf V}_{\rm sep} \right| \la 20$ m s$^{-1}$, consistent with split events of other comets \citep[][and citations therein]{2004come.book..301B}. Since the observed arc of the two fragments is insufficiently long to allow for a robust determination of the split parameters as wholly free parameters due to the existence of multiple local minima in the goodness of fit, we varied the split epoch $t_{\rm frg}$ with an incremental step size of 10$^{-3}$ d and solved for the RTN components of the separation velocity as free parameters. The results centred around the time of perihelion passage of the primary are plotted in Figure \ref{fig:bestfit_frg}, together with the dimensionless goodness of fit $\chi_{\rm frg}^{2}$. What we found is that, although the minima in the goodness of fit of the two fragments both lie within $\sim$6 hr before the perihelion passage of the primary, the temporal trends are dissimilar to each other, as Fragment A clearly possesses a well-defined dip with $\min \left\{\chi_{\rm frg}^{2} \right\} \approx 4$, whereas there is no easily noticeable dip but a bump for Fragment B. If the fragments split from the primary much earlier than the main mass-shedding event near perihelion, the goodness of fit is comparable to yet marginally greater than $\min \left\{\chi_{\rm frg}^{2} \right\}$ near perihelion. Postperihelion solutions for Fragment B with split epochs in late 2021 February and slightly worse goodness of fit also exist, but once again there is no easily recognisable dip in $\chi_{\rm frg}^{2}$. Therefore, we cannot constrain the split epoch of Fragment B except that it was less likely to come off from the primary around $\sim$4 hr before the perihelion passage of the primary.

In the event that Fragment A split from the primary around the perihelion passage of the latter on 2021 January 17, we will then be able to robustly constrain the split parameters. We treated the split epoch as an additional free parameter. Regardless of what initial guess for the split epoch, as long as it is around the dip in the goodness of fit, and what initial guesses for the components of the separation velocity were tried, our code always swiftly converged to the identical solution arriving at the smallest goodness of fit (Solution II for Fragment A in Table \ref{tab:fit_frg}). 

We also tested the reliability of the reported formal errors on the split parameters in Solution II for Fragment A in Table \ref{tab:orb_frg} by repeatedly solving for the split parameters with the 300 MC variant orbits created in Section \ref{ssec_orbevo}, rather than the nominal one. Our result is that the standard deviations are completely consistent with the $1\sigma$ formal errors propagated from uncertainties in our astrometric measurements. However, we note that Solution II for Fragment A has an excessively large separation speed, basically in the out-of-plane direction. Separation speeds of similarly enormous are rarely seen for split comets, including other near-Sun comets such as Kreutz-family sungrazers and Marsden- and Kracht-group sunskirters \citep[e.g.,][]{2004ApJ...607..620S,2005ApJS..161..551S}, but are not unheard of \citep[][and citations therein]{2004come.book..301B}. An attempt we have made was to include the radial and/or transverse nongravitational parameters as an additional parameter in our fragmentation model. However, this resulted in no improvement for Fragment A whatsoever. We also briefly explored in the same fashion with trial split epochs around the previous perihelion passage of 323P on 2016 November 23, finding that our astrometric observations would permit the existence of such a solution, in which the separation velocity of Fragment A lies completely in the orbital plane of the primary and the split epoch is $\sim$1.4 d before the perihelion (Solution III for Fragment A in Table \ref{tab:orb_frg}). 

To conclude, we were not able to find unique solutions for the split parameters of the two fragments. If the fragmentation event occurred around the perihelion passage in January 2021, Fragment A would split from the primary at an uncommonly huge separation speed, while the separation speed of Fragment B would be much lower, similar to those of other split comets. It is possible that Fragment A did not form during the perihelion passage in 2021 but from some earlier epoch such as the previous perihelion return. The major impediment to our attempts is primarily attributed to accumulation of errors from propagating the short observed arcs of the fragments.

\begin{deluxetable*}{lc|c|c|c|c|c}
\tablecaption{Selected Best-Fit Fragmentation Solutions
\label{tab:fit_frg}}
\tablewidth{0pt}
\tablehead{
&  & 
\multicolumn{3}{c|}{Fragment A}  & 
\multicolumn{2}{c}{FragmentB} \\ \cline{3-7}
\multicolumn{2}{c|}{Quantity} &
Solution I & 
Solution II & Solution III
&
Solution I & Solution II
}
\startdata
Split epoch (TDB)\tablenotemark{$\dagger$} & $t_{\rm frg}$
       & 2020 Feb 13.0
       & 2021 Jan $17.5190 \pm 0.0031$
       & 2016 Nov $22.267 \pm 0.094$
       & 2021 Jan 17.4
       & 2021 Feb 24.4 \\ 
Separation velocity (m s$^{-1}$) & & & & & \\
~~~~Radial component & $V_{\rm R}$
       &  $-0.228 \pm 0.014$ 
       & $+2.12 \pm 0.24$
       & $+0.47 \pm 0.11$
       & $+2.11 \pm 0.21$ 
       & $+6.09 \pm 0.97$\\ 
~~~~Transverse component & $V_{\rm T}$
       & $-0.015 \pm 0.012$ 
       & $+2.255 \pm 0.097$
       & $-0.232 \pm 0.091$
       & $+0.46 \pm 0.15$
       & $-6.2 \pm 1.2$\\ 
~~~~Normal component & $V_{\rm N}$ 
       & $\left(+9.40 \pm 0.51\right) \times 10^{-3}$
       & $+12.1 \pm 2.5$
       & 0
       & $-1.57 \pm 0.14$
       & $-0.235 \pm 0.051$\\  \hline
Goodness of fit & $\chi_{\rm frg}^{2}$
                 & 4.585 
                 & 
                 4.361
                 & 4.448
                 & 7.020
                 & 7.033\\
Normalised RMS residuals & 
& 0.505
& 
0.492
& 0.497
& 0.624
& 0.625\\
\enddata
\tablenotetext{\dagger}{The corresponding uncertainty is in days.}
\tablecomments{Both the goodness of fit and normalised rms residuals are dimensionless. All of the reported uncertainties are $1\sigma$ formal errors propagated from the astrometric measurement uncertainties of the two fragments and the primary. Solutions having split parameters without uncertainties given mean that the corresponding parameters were held fixed therein. We are aware that although Solution II for Fragment A has the best goodness of fit in the timespan we examined, it renders us an excessively large normal component of the separation velocity that is rarely seen but by no means unheard of for split comets \citep[][and citations therein]{2004come.book..301B}.}
\end{deluxetable*}

\subsection{Mass-Loss Mechanism}
\label{ssec_mlm}

First, we estimate the mass of the postperihelion dust ejecta using our aperture photometric measurements. Given the size distribution of the dust ejecta, while its effective scattering cross-section is dominated by the smallest dust, the mass is not. The mass ratio between the dust ejecta and the nucleus can be calculated from
\begin{equation}
\zeta = \left(\frac{\gamma - 3}{4 - \gamma} \right) \left(\frac{\mathfrak{a}_{\rm c}}{\mathfrak{a}_{\max}}\right)^{\gamma - 3} \left(\frac{\rho_{\rm d} \mathfrak{a}_{\max}}{\rho_{\rm n} R_{\rm n}} \right) \eta
\label{eq_mratio}.
\end{equation}
\noindent Substitution accordingly into the above equation yields that the mass ratio between the dust ejecta enclosed by our largest aperture and the nucleus is $\sim$0.1-10\%. Given the fact that the whole dust ejecta of the comet evidently extended far beyond the aperture, we regard the obtained value as a lower limit only. If 323P sheds mass similarly whenever around perihelion, it is expected to disappear after $\la$10-10$^3$ revolutions around the Sun, or within a timescale of no more than a few millennia.

Despite that 323P has an unexceptional cometary designation assigned by the Minor Planet Center, we argue that it is distinct from typical comets in the solar system, whose main activity is driven by sublimation of volatiles, because the temperature is too high for water ice, the typical dominant cometary volatile \citep[e.g.,][]{1950ApJ...111..375W}, to survive. To see this, we calculate the thermal timescale, on which solar heat received at the surface propagates towards the deep interior of the nucleus as $\tau_{\rm th} = R_{\rm n}^{2} / \left(\kappa \pi^2\right)$, where $\kappa \sim 10^{-7}$-10$^{-6}$ m$^{2}$ s$^{-1}$ is the thermal diffusivity for typical rocks \citep{2004come.book..359P}. Inserting the nucleus size we obtained in Section \ref{ssec_xs}, we find $\tau_{\rm th} \sim 20$-200 yr, which is much shorter than the age of 323P since it became a sunskirter ($\sim$1 kyr, see Section \ref{ssec_orbevo}) and is also orders of magnitude shorter than the dynamical lifetimes of near-Earth objects in planet-crossing orbits \citep[$\sim$1-10 Myr;][]{2000Icar..146..176G, 2002Icar..159..358F}. Consequently, we expect that the interior of the nucleus of 323P has been warmed up to a degree where the survival of water ice is highly dubious due to heat conduction from the surface since the commencement of its sunskirting dynamical status.

Next, we estimate the core temperature of the nucleus of the comet from the energy conservation law
\begin{align}
\nonumber
T_{\rm c} & = \frac{1}{2}\left[\frac{\left(1 - A_{\rm B} \right) L_{\odot}}{\pi \epsilon \sigma P} \int\limits_{t_{0}}^{t_{0} + P} \frac{{\rm d} t}{r_{\rm H}^2 \left(t\right)} \right]^{1/4} \\
& \approx \frac{1}{2} \left[\frac{\left(1 - A_{\rm B}\right) L_{\odot}}{\pi \epsilon \sigma a^2 \sqrt{1 - e^2}} \right]^{1/4}
\label{eq_Tc},
\end{align}
\noindent where $A_{\rm B}$ and $\epsilon$ are respectively the Bond albedo and emissivity of the nucleus, $\sigma = 5.67 \times 10^{-8}$ W m$^{-2}$ K$^{-4}$ is the Stefan-Boltzmann constant, and $t_{0}$ is some arbitrary referenced epoch. Given the typical ranges of the emissivity and Bond albedo for asteroids and comets in the solar system \citep[see, e.g.,][]{2004come.book..223L,2005Icar..173..153L,2016ApJ...817L..22L}, we find $T_{\rm c} \approx 260$-270 K for the core temperature of the cometary nucleus, which is unpromising for the nucleus of 323P to maintain water ice within its dynamical lifetime.

We proceed to evaluate the survival of water ice in the nucleus of 323P in the order of magnitude calculation manner by assuming that the whole nucleus is solely comprised of water ice undergoing free sublimation as the Sun heats the nucleus surface. As a result, after an amount of time $\Delta t$, the nucleus shrinks by $\left| \Delta R_{\rm n} \right|$ in radius, corresponding to a mass loss of $\sim\!4\pi \rho_{\rm n} R_{\rm n}^{2} \Delta R_{\rm n}$ attributed to the sublimation activity. We estimate the timescale for such a sublimating water-ice nucleus to exist from the energy conservation law as
\begin{align}
\nonumber
\tau_{\rm ice} & \equiv \frac{R_{\rm n}}{\left| \overline{\dot{R}_{\rm n}} \right|} \\
& \approx \frac{4 \pi \chi \rho_{\rm n} R_{\rm n} \mathscr{L} a^{2} \sqrt{1 - e^2}}{\left(1 - A_{\rm B} \right) L_{\odot}}
\label{eq_tau_ice}.
\end{align}
\noindent Here, $\chi$ is the adimensional illumination coefficient in a range from 1 to 4, respectively corresponding to the subsolar and isothermal scenarios, and $\mathscr{L} \approx 3 \times 10^{6}$ J kg$^{-1}$ is the latent heat for water ice to sublimate. Substituting, we obtain that such an icy nucleus would have been annihilated within a timescale as short as $\tau_{\rm ice} \la 10^{2}$ yr, which is once again much shorter than the age of 323P since it became a sunskirter. It is needless to mention more volatile substances, such as carbon monoxide and carbon dioxide. Accordingly, we can conclude that even though 323P may have once contained icy ingredients as an ordinary Jupiter-family comet before the current sunskirting dynamical status, it is highly unlikely that its current nucleus still maintains any volatile components, and we can confidently negate the possibility that the observed mass loss of 323P is still driven by sublimation of volatiles. In this regard, 323P is readily distinct from typical comets in the solar system.

As mentioned in Section \ref{ssec_rot}, the rotation period we found for the nucleus of the comet, $P_{\rm rot} \approx 0.522$ hr, is the shortest amongst known comets in the solar system. We argue that the nucleus could be spun up by the Yarkovsky-O'Keefe-Radzievskii-Paddack (YORP) effect after becoming a dead comet (if at all possible), because its YORP timescale, estimated from the relation derived by \citet{2015aste.book..221J}, is merely $\sim$10 kyr, much shorter than typical dynamical lifetimes for near-Earth objects. From Equation (\ref{eq_cs}), any internal cohesive strength $\la$10-100 Pa within the interior will place the nucleus in the regime of rotational instability, whereby mass shedding will take place. Therefore, it seems totally plausible that the nucleus of 323P is rotationally disrupting. 

Besides, based on the fact that 323P showed signs of activity around perihelion, we also consider the possibility of the observed mass loss of the comet being triggered by thermal fracture, as in the case of near-Sun asteroid (3200) Phaethon \citep{2010AJ....140.1519J}. Here we assess the thermal stress arising from any temperature gradient across the nucleus of the comet in the order-of-magnitude fashion, as detailed computation is beyond the scope of this paper. At perihelion, the nucleus surface reaches an equilibrium temperature of $\ga$1400 K. Thus, a temperature gradient of $\Delta T \ga 1000$ K across the whole nucleus of the comet is expected. Assuming an elastic scenario, the resulting thermal stress can be then approximated by $\sigma_{\rm th} \sim \alpha_{\rm V} E_{\rm Y} \Delta T$, where $\alpha_{\rm V} \sim 10^{-5}$ K$^{-1}$ is the volumetric thermal expansion coefficient and $E_{\rm Y} \sim 10$-100 GPa is the Young's modulus typical for rocks \citep[e.g.,][]{2015JGRE..120..255M}. Substituting, we find that the nucleus of 323P suffers from an enormous thermal stress of $\sigma_{\rm th} \sim 0.1$-1 GPa within its interior around its perihelion passages, not only orders of magnitude larger than the cohesive strengths of cometary nuclei and asteroids, but also larger than the tensile strengths of meteorites \citep[e.g.,][and citations therein]{2015aste.book..745S,2019SSRv..215...29G}. Thus, thermal fracturing is expected to occur to the nucleus of 323P near perihelion, during which a portion of the nucleus cracks and crumbles, producing postperihelion dust ejecta.

\section{Summary}
\label{sec_sum}

Our observing campaign with a mix of ground and space telescopes successfully recorded the mass loss of 323P/SOHO in great detail for the very first time for a periodic near-Sun object. The key conclusions of our study are:
\begin{enumerate}

\item While the comet was not noticeably active in the preperihelion Subaru observation from December 2020, it possessed an obvious tail of a few arcminutes in length mimicking a disintegrated comet postperihelion from February to March 2021. The total mass of the ejecta is estimated to be at least 0.1-10\% of the nucleus mass.

\item Our syndyne-synchrone computation indicates that the main dust ejecta, consisting of at least millimetre-sized dust grains, was formed within a day past the perihelion passage in an impulsive manner. However, the existence of its antitail around the plane-crossing time suggests older mass shedding of decimetre-size or larger boulders from the nucleus no later than November or December 2020. Our aperture photometry suggests the dust ejecta following a size distribution with power-law index $\gamma = 3.2 \pm 0.2$.

\item Two fragments of the comet, both of $\sim$20 m in radius, were observed in the first two visits of the HST observations. Unfortunately, we cannot robustly determine their separation velocities and splitting epochs from the primary, because they were undetected in subsequent HST observations from late March 2021. 

\item The nucleus of the comet, with an effective radius of $86 \pm 3$ m (assuming a geometric albedo of 0.15) and an aspect ratio of $\sim$0.7, is likely in the state of rotational instability, as its rotation period, $P_{\rm rot} \approx 0.522$ hr, is the shortest for known comets in the solar system. We accordingly estimated the cohesive strength of its interior to be $\ga$10-100 Pa.

\item We observed that the postperihelion colour of the nucleus noticeably reddened in the $g - r$ regime but basically remained unchanged in $r - i$. The mean colour indices of the nucleus were $g - r = 0.44 \pm 0.06$ and $r - i = -0.08 \pm 0.09$ on 2021 February 13, and $g - r = 0.68 \pm 0.05$ and $r - i = -0.16 \pm 0.05$ on March 3. We also measured the mean colour indices of the dust ejecta to be $g - r = 0.85 \pm 0.07$ and $r - i = -0.07 \pm 0.11$ on February 13, and $g - r = 0.37 \pm 0.05$ and $r - i = -0.11 \pm 0.07$ on March 3. These colours are freakish compared to those of other small solar system bodies. The way that the colours of the nucleus and the ejecta temporally varied has never been witnessed before amongst other small solar system bodies.

\item From the astrometric measurements of the comet, we found a statistically confident transverse component of the nongravitational parameter $A_{2} = \left(-6.499 \pm 0.009 \right) \times 10^{-20}$ au d$^{-2}$, with the nongravitational acceleration following a steep heliocentric dependency of $r_{\rm H}^{-8.5}$.

\item The comet has a likelihood of 99.7\% to collide with the Sun within the next two millennia owing to the strong $\nu_{6}$ secular resonance, which effectively shrinks its perihelion distance while increasing the eccentricity.

\item While the comet may have been an ordinary Jupiter-family comet $\ga$1 kyr ago, its observed mass loss cannot be explained by sublimation of cometary volatiles. Rather, it is most likely triggered by the rotational instability, plus the enormous thermal stress induced by the huge temperature gradient within its nucleus around perihelion.

\end{enumerate}

\begin{acknowledgements}
We thank the anonymous reviewer, Jian-Yang Li, and Matthew M. Knight for insightful comments, Bill Gray and Aldo Vitagliano for implementing {\tt FindOrb} and {\tt SOLEX12}, respectively, and Karl Battams for providing us with his SOHO astrometry of 323P. MTH thanks great help from Josef {\v D}urech and Jon D. Giorgini and support from Kiwi. DJT thanks queue observers C. Crowder, C. Cunningham, H. Januszewski, L. Wells, and C. Wipper at CFHT for obtaining the CFHT observing data, and support from NASA grant 80NSSC21K0807. PAW acknowledges support from the Natural Sciences and Engineering Research Council of Canada Discovery Grants program (Grant no. RGPIN-2018-05659). This research is based on on data obtained at the Canada-France-Hawaii Telescope (CFHT), the international Gemini Observatory, the NASA/ESA Hubble Space Telescope obtained from the Space Telescope Science Institute, the Lowell Discovery Telescope (LDT) at Lowell Observatory, and the Subaru Telescope. MegaPrime/MegaCam at CFHT is a joint project of CFHT and CEA/DAPNIA, operated by the National Research Council (NRC) of Canada, the Institut National des Science de l'Univers of the Centre National de la Recherche Scientifique (CNRS) of France, and the University of Hawaii. The international Gemini Observatory is a program of NSF's NOIRLab managed by the Association of Universities for Research in Astronomy (AURA) under a cooperative agreement with the National Science Foundation on behalf of the Gemini Observatory partnership: the National Science Foundation (United States), National Research Council (Canada), Agencia Nacional de Investigaci\'{o}n y Desarrollo (Chile), Ministerio de Ciencia, Tecnolog\'{i}a e Innovaci\'{o}n (Argentina), Minist\'{e}rio da Ci\^{e}ncia, Tecnologia, Inova\c{c}\~{o}es e Comunica\c{c}\~{o}es (Brazil), and Korea Astronomy and Space Science Institute (Republic of Korea). The NASA/ESA Hubble Space Telescope is operated by the Association of Universities for Research in Astronomy, Inc., under NASA contract NAS 5-26555. Lowell is a private, non-profit institution dedicated to astrophysical research and public appreciation of astronomy and operates the LDT in partnership with Boston University, the University of Maryland, the University of Toledo, Northern Arizona University and Yale University. The Large Monolithic Imager at the LDT was built by Lowell Observatory using funds provided by the National Science Foundation (AST-1005313). The Subaru Telescope is operated by the National Astronomical Observatory of Japan. We are honoured and grateful for the opportunity of observing the Universe from Maunakea, which has the cultural, historical, and natural significance in Hawaii.
\end{acknowledgements}

\vspace{5mm}
\facilities{2.4 m HST, 3.6 m CFHT, 4.3 m LDT, 8.1 m Gemini North, 8.2 m Subaru}

\software{{\tt FindOrb}, lightcurve inversion software package \citep{2010A&A...513A..46D}, {\tt MPFIT} \citep{2009ASPC..411..251M}, 
{\tt SOLEX12} \citep{1997CeMDA..66..293V}, {\tt TinyTim} \citep{2011SPIE.8127E..0JK}}




\end{document}